\documentclass[aps,prd,superscriptaddress,floatfix,nofootinbib,notitlepage]{revtex4-1}

\usepackage{aas_macros}
\usepackage{graphicx}
\usepackage{multirow}
\usepackage{amsmath,amssymb,amsfonts}
\usepackage{verbatim}
\usepackage{bm}
\usepackage{color}
\usepackage{ulem}
\usepackage{mathtools}
\usepackage{colortbl}
\usepackage[dvipsnames]{xcolor}

\usepackage[colorlinks]{hyperref}
\hypersetup{
    colorlinks=true,
    linkcolor=blue,
    filecolor=blue,      
    urlcolor=blue,
    citecolor=magenta,
    pdfpagemode=FullScreen
}

\renewcommand{\vec}[1]{\bm{#1}}
\newcommand{\orcid}[1]{\hspace{1mm}\href{https://orcid.org/#1}{\includegraphics[height=0.3cm,keepaspectratio]{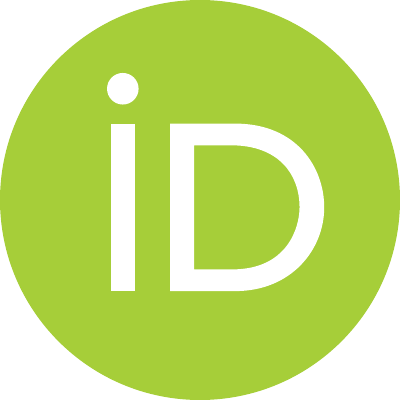}}}

\begin{document}

\title{Galactic factories of cosmic-ray electrons and positrons}

\author{Carmelo Evoli\orcid{0000-0002-6023-5253}} 
\email{carmelo.evoli@gssi.it}
\affiliation{Gran Sasso Science Institute (GSSI), Viale Francesco Crispi 7, 67100 L'Aquila, Italy}
\affiliation{INFN-Laboratori Nazionali del Gran Sasso (LNGS),  via G. Acitelli 22, 67100 Assergi (AQ), Italy}

\author{Elena Amato\orcid{0000-0002-9881-8112}}
\email{elena.amato@inaf.it}
\affiliation{INAF-Osservatorio Astrofisico di Arcetri, Largo E. Fermi 5, 50125 Firenze, Italy}

\author{Pasquale Blasi\orcid{0000-0003-2480-599X}}
\email{pasquale.blasi@gssi.it}
\affiliation{Gran Sasso Science Institute (GSSI), Viale Francesco Crispi 7, 67100 L'Aquila, Italy}
\affiliation{INFN-Laboratori Nazionali del Gran Sasso (LNGS),  via G. Acitelli 22, 67100 Assergi (AQ), Italy}

\author{Roberto Aloisio\orcid{0000-0003-0161-5923}}
\email{roberto.aloisio@gssi.it}
\affiliation{Gran Sasso Science Institute (GSSI), Viale Francesco Crispi 7, 67100 L'Aquila, Italy}
\affiliation{INFN-Laboratori Nazionali del Gran Sasso (LNGS),  via G. Acitelli 22, 67100 Assergi (AQ), Italy}

\begin{abstract}
We present a novel calculation of the spectrum of electrons and positrons from random sources, supernova remnants and pulsars, distributed within the spiral arms of the Galaxy. The pulsar emissivity in terms of electron-positron pairs is considered as time dependent, following the magnetic dipole spin-down luminosity, and the temporal evolution of the potential drop is accounted for. Moreover each pulsar, with the magnetic field and initial spin period selected at random from the observed distribution, is considered as a source of pairs only after it leaves the parent supernova due to its birth kick velocity (also selected at random from the observed distribution). We show that (i) the spectrum of electrons is characterized by a feature at $\gtrsim 50$ GeV that proves that their transport is dominated by radiative losses. The flux reduction at $E\gtrsim 1$ TeV is explained as a result of lepton transport from sources in the spiral arms. (ii) The spectrum of positrons is very well described by the contribution of pulsars and the rising positron fraction originates naturally. The implications of pulsars as positron sources in terms of positron fraction at very high energies are also discussed. (iii) The role of fluctuations in the high-energy regime is thoroughly discussed and used to draw conclusions on the possibility to single out the contribution of local sources to the lepton spectrum with current and upcoming experiments. 
\end{abstract}

\maketitle

\section{Introduction}

Electrons and positrons in the cosmic radiation are important pieces of a puzzle, that of the origin of cosmic rays (CRs), that need to fall in the right places, together with the other numerous pieces associated with both primary and secondary nuclei~\cite{Blasi2013aarv,AmatoBlasi2018,Gabici2019rev}. The consensus picture that has recently arisen based on AMS-02 measurements of the fluxes of primary nuclei~\cite{AMS02.2015.protons,AMS02.2017.heco} is that the transport of these particles is dominated by diffusion and advection (the latter being important at rigidity $\lesssim 10$ GV) in combination with the unavoidable spallation losses, which in turn give rise to the production of secondary nuclei, such as boron, lithium and beryllium~\cite{AMS02.2018.libeb}. The measurement of the secondary to primary ratios has impacted the field in a substantial way, by (i) providing evidence for the fact that the spectral break observed in primary nuclei at $\sim 200-300$ GV is even more pronounced in secondary nuclei, hence confirming that the break has to be associated with a corresponding change of slope in the diffusion coefficient that CRs experience while propagating through the Galaxy; and (ii) providing a better evaluation of the grammage traversed by CRs in the Galaxy, both in terms of normalization and rigidity dependence. 

This picture would seem well established if it were not for two important anomalies. One, the most prominent, is the excess in the positrons as measured at Earth: in the standard picture of the origin of CRs, positrons can only result from the decay of charged pions produced in the inelastic collisions of CR protons and helium nuclei with the interstellar medium (ISM)~\cite{Protheroe1982}, and their spectrum is steeper than both primary protons and even secondary nuclei such as boron because of the effect of radiative losses. Within this scenario, the ratio between the flux of positrons and that of electrons+positrons is a decreasing function of energy~\cite{Moskalenko1998apj}. Early measurements of this so-called positron fraction yielded a preliminary and yet puzzling evidence for a flat or even growing trend~\cite{Muller1990icrc,HEAT.1995.posfraction,AMS01.2007.posfraction}, at odds with the standard model of CR transport. But the statistics of these early measurements, due to the very low fluxes of positrons in the cosmic radiation, was too poor to lead to strong conclusions. However this finding found a spectacular confirmation in the results of the PAMELA experiment~\cite{PAMELA.2009.posfraction}, showing a positron fraction growing with energy at least up to $\sim 100$ GeV. The excess was also found by Fermi-LAT using the magnetic field of Earth to discriminate electrons from positrons~\cite{FERMI.2017.leptons}. This strong anomaly led to many possible explanations based on new physics, associated with dark matter annihilation in the Galaxy~(see~\cite{Serpico2012aph} and references therein). 
The positron excess was later confirmed and extended to even larger energies and measured with higher accuracy by the AMS-02 experiment on board the International Space Station~\cite{AMS02.2013.posfraction}. The unambiguous measurement of the spectra of electrons~\cite{AMS02.2019.electrons} and positrons~\cite{AMS02.2019.positrons} separately demonstrated that the rise in the positron fraction is due to an excess of positrons rather than a deficit of electrons. The spectra of electrons and positrons are clearly different, and the spectrum of positrons is disturbingly close to that of protons. 

The second anomaly is associated with the ratio of the antiproton to proton flux as measured by AMS-02~\cite{AMS02.2016.pbar}. The ratio is rather flat with energy while it is expected to decrease to an extent that depends on both the diffusion coefficient experienced by the parent protons and the daughter antiprotons and to the cross section for antiproton production, which increases with energy. This second anomaly is likely to reflect more our uncertainties on the cross sections rather than a physical problem~\cite{Boudaud2020prr}.

Nevertheless, the combined finding of a hard positron spectrum and a hard antiproton spectrum led some authors to question the pillars of the standard model of CR transport, while still assuming that both positrons and antiprotons are purely secondary products of CR interactions~\cite{Blum2013prl,Cowsik2016apj,Lipari2017prd,Kachelriess2018prd}. 
As a by-product of this proposal, one would be forced to require a small Galactic halo for CR diffusion and negligible radiative energy losses for electrons and positrons, at least up to $\sim 1$ TeV. The recent measurement of the beryllium spectrum with AMS-02, and especially the ratio of beryllium to boron fluxes~\cite{AMS02.2018.libeb}, has been used to infer a lower limit to the size $H$ of the halo to $H\gtrsim 5$ kpc~\cite{Evoli2020prd,Weinrich2020aaa}. Such a large halo is difficult to reconcile with the assumption of negligible energy losses for leptons.
Moreover in a recent article~\cite{Evoli2020prl} we have discussed the possibility that the feature at $E\gtrsim 50$ GeV in the electron spectrum is due to the onset of Klein-Nishina (KN) effects in the inverse Compton scattering (ICS) of these particles on the UV background light of the Galaxy, as also discussed in some previous work~\cite{Aharonian1985afz,derWalt1991mnras,Stawarz2010apj,Schlickeiser2010njph}. This finding might prove that electrons' transport is dominated by radiative losses on Galactic scales, consistent with the standard model of CR transport. Hence a source of positrons other than CR interactions in the ISM is needed in order to accommodate the rising positron fraction and even the positron spectrum itself; see also~\cite{Diesing2020prd}. 

Pulsars were proposed as sources of positrons long before the PAMELA discovery of the positron excess ~\cite{Harding1987icrc,Boulares1989apj,Atoyan1995prd}, but studies of their role as positron sources proliferated immediately after. It was soon recognized that data could be described very naturally in terms of the positron contribution from old neutron stars~\cite{Hooper2009jcap}, although the issues of which type of pulsars and escape of pairs from the magnetosphere remained unanswered. Several other studies followed~\cite{Grasso2009aph,Yuksel2009prl,Serpico2009prd,Delahaye2010aa,Profumo2012,Manconi2017jcap} also investigating the role of local sources. In Ref.~\cite{BlasiAmato2011} it was proposed that the main contributors be middle-aged neutron stars produced in core collapse supernova explosions with large birth kick velocities, which would lead them to escape the parent remnant within a few $\times 10^4$ yr after the explosion. The vast majority of pulsars are indeed expected to move supersonically with respect to the local ISM and generate a bow shock nebula (BSN) in the ISM~\citep{Bykov2017ssrv}; then the pairs may be able to leave the magnetosphere from the tail of the nebula. The fact that $e^{-}+e^{+}$ pairs can escape the parent pulsar environment has recently been confirmed by the detection by the HAWC~\cite{HAWC.2017.tevhalos} and MILAGRO~\cite{MILAGRO.2009.tevsources} observatories of extended diffuse gamma-ray emission, apparently due to ICS of extremely high-energy leptons around selected pulsars~\cite{HAWC.2017.tevhalos}, all old and outside their parent supernova remnant (SNR)~\cite{Sudoh2019prd,Giacinti2020aa}. The presence of a $\gamma$-ray halo around Geminga has been confirmed with an analysis of Fermi-LAT data by~\cite{DiMauro2019prd}.

In recent years there has also been a dramatic improvement in the measurement of the spectrum of $e^{-}+e^{+}$ (hereafter leptons) at high energies. HESS reported the first measurement of the lepton spectrum up to 5 TeV, showing substantial steepening at $\sim 900$~GeV ($\Delta \gamma \sim 1$)~\cite{HESS.2008.leptons,HESS.2009.leptons}. 
CALET~\cite{CALET.2017.leptons} and DAMPE~\cite{DAMPE.2017.leptons} have provided the first direct measurements of the total lepton spectrum up to $\sim 5$~TeV. DAMPE largely confirmed the spectral softening at about 0.9 TeV with the spectral index changing from $\sim 3.1$ to $\sim 3.9$. Such softening is most likely to be attributed to transport from nearby sources rather than to a cutoff in the source spectrum~\cite{Kobayashi2004apj,Mertsch2011jcap}.

The spectrum of leptons is dominated by the contribution of primary electrons, presumably accelerated by the same sources responsible for the spectrum of CR nuclei. Particle in cell simulations confirm that collisionless shocks accelerate electrons although with a smaller efficiency (by about a factor $\sim 50$) than protons~\cite{Jaehong2015prl}, consistent with the ratio of electrons to protons in CRs. Diffusive shock acceleration (DSA) is not sensitive to the charge; hence, the slope of the instantaneous spectrum of accelerated electrons is the same as for protons. Yet, phenomenological approaches to the transport of electrons in the Galaxy imply an injection spectrum at sources slightly steeper than that of protons~\cite{DiBernardo2013jcap}. This difference might be accounted for as due to radiative energy losses inside the sources, most likely SNRs~\cite{Diesing2019prl}, if the magnetic field in the late stages of the evolution is large enough. The acceleration of electrons at supernova shocks is better established than the acceleration of hadrons at the same shocks because of the larger radiative efficiency of electrons. Not only SNRs have long been known as sources of synchrotron emission at radio wavelengths, but more recently nonthermal x-ray emission has been detected from virtually all young SNRs~\cite{Vink2012aarv}, and the morphology of the emission has provided us with the best evidence so far of magnetic field amplification at SNR shocks and of electron acceleration up to $\gtrsim 10$ TeV energies. Finally, gamma-ray emission from several SNRs has now been attributed to radiative losses of electrons (see \cite{Funk2017book} and references therein for a recent review), mostly resulting in hard gamma-ray spectra, as expected for ICS photons. 

In the present article we present the state of the art assessment of the role of SNRs and pulsar wind nebulae (PWNe) for the production of CR electrons and positrons in a structured Galaxy consisting of spiral arms, where sources are located, and ISM, where interactions of CR nuclei give rise to secondary leptons. The calculation is based on a Green function formalism with proper boundary conditions that allows us to (i) generate SN events with a Monte Carlo technique and follow electrons from the sources to Earth (as previously done, e.g.,~in~\cite{Delahaye2010aa,Mertsch2011jcap,Cholis2018prd,Manconi2020prd}); (ii) generate the locations of pulsars, with their initial period and kick velocity, also chosen at random according to the measured distributions, and determine the time in which the pulsar escapes the parent SNR and gives rise to a BSN. We assume that this is the phase when pairs are released into the ISM and the temporal evolution of the injection is followed accordingly, assuming a magnetic dipole spin down. The initial period of the pulsar and the kick velocity is generated from distributions consistent with the observed statistics of objects. (iii) calculate the spectrum of electrons and positrons generated as secondary products of CR interactions with the ISM gas, also retaining the $e^{+}$-$e^{-}$ asymmetry.   

Previous attempts employed mixed approaches where the sources of leptons are separated in two classes, namely that of distant sources, treated as a continuum, and local sources~\cite{Kachelriess2015prl,Recchia2019prd,Fornieri2020jcap}, eventually picked from catalogs of either SNRs or pulsars or both~\cite{DiBernardo2011aph,Manconi2019jcap}. Notice that the choice of using catalogs for the local sources may not be well justified for several reasons. First, the catalogs are generally incomplete; second, the fact of having a nearby source does not necessarily mean that it provides a large contribution to the flux of leptons at Earth. For instance the magnetic field around the source or around Earth might lead the particles to move in directions that are unfavorable to reaching us. A crucial role in assessing this point would be played by the knowledge of the strength and coherence scale of the field, and by the induced anisotropy in the diffusive processes in the directions parallel and perpendicular to the field. Third, the separation between what are selected as local sources and what are labeled as uniformly distributed sources is somewhat arbitrary and catalog related rather than being based on physical motivations. This is an especially delicate issue in the presence of spiral arms. For these reasons we claim that it is preferable for this type of calculations to rely on Monte Carlo techniques, as done here and recently in~\cite{Mertsch2011jcap}, where however the attention was focused on electrons from SNRs only. This choice is very challenging in terms of computation time, since a large number of sources need to be followed in time (due to time-dependent injection in the case of pulsars) on timescales of the order of 100 million years, so as to be sure that the stationary regime is reached and transients are avoided. 

The other main novelties and results of the calculations can be summarized as follows: 
(a) We include, for the first time, the time dependence of the injection of pairs by pulsars, and we account for the fact that pairs are only released during the BSN phase, starting when the pulsar leaves the parent SNR. A proper discussion of the differences between impulsive release and time-dependent injection is presented. A previous attempt to implement the time dependence of the injection was presented in~\cite{Cholis2018prd}, where however a number of simplifying assumptions were made: the spiral distribution of sources was not included and Inverse Compton losses were only treated in the Thomson approximation.
(b) We find that, in order to reproduce the positron spectrum in the AMS-02 energy range, the spectrum of pairs released by PWNe must be a bit softer than typically inferred from multiwavelength observations of these sources. At the same time, the electrons produced by SNRs must be injected with a spectrum that is softer (with a slope different by $\sim 0.3$) than that of protons.
At energies higher than those measured by AMS-02, the total spectrum of leptons, including the contribution from SNRs, pulsars and CR interactions, exhibits a steepening. This results from the combination of the intrinsic cutoff in the SNR injection spectrum and the losses suffered by the particles during galactic transport.~The computed electron+positron spectrum appears to be fully compatible with the available measurements.
(c) We show that the positron fraction is also very well described and discuss the predicted trend of the same quantity for future experiments that could extend measurements to higher energies. 
(d) The role of fluctuations and the corresponding expected cosmic variance is discussed and quantified for all the considered scenarios. These effects become important at high energies, $\gtrsim 10$ TeV. Indeed, we show that the number of sources contributing to the local spectrum is significantly larger than found in previous works, and it becomes of $\mathcal{O}(10)$ only at energies $\gtrsim 10$~TeV.
(e) We find the spiral arms to be of the greatest importance in that most sources are located at a distance from the Sun that is the mean distance to the closest arm: this leads to an electron spectrum that is overall slightly steeper than predicted in the absence of arms and to a smaller role of cosmic variance. The effect of uncertainties in the position of the Sun in the Galaxy is also discussed. 
(f) Given the effort of some ongoing experiments, such as CALET and DAMPE, aimed at the identification of local sources in the lepton spectrum, we make an attempt to quantify the spectral fluctuations of leptons at high energies by showing how the spectrum measured at Earth changes for a few selected realizations.

Finally, we implement in our calculations the very recent parametrization provided by~\cite{Fang2020arxiv} which describes with great accuracy the transition to the KN regime in the energy loss rates. In Sec.~\ref{sec:green} we inspect the impact of the new parametrization on the possibility to identify the feature observed by AMS-02 in the electron spectrum with the onset of KN effects on the ICS scattering off UV light as found in~\cite{Evoli2020prl}.

The article is structured as follows: in Sec.~\ref{sec:green} we summarize the formalism based on Green functions constrained by proper boundary conditions and we recall crucial aspects of lepton transport in the presence of radiative losses. In Sec.~\ref{sec:leptonsources} we discuss the different sources of electrons and positrons in the Galaxy. In Sec.~\ref{sec:spirals} we describe in detail the spatial distribution of sources in the spiral structure of our Galaxy and the way we generate Monte Carlo realizations of SNRs and pulsars. The results of our work are presented in Sec.~\ref{sec:results}, while our conclusions are drawn in Sec.~\ref{sec:conclusions}.

\section{Green functions formalism}
\label{sec:green}

\begin{figure*}[t]
\centering
\hspace{\stretch{1}}
\includegraphics[width=0.48\textwidth]{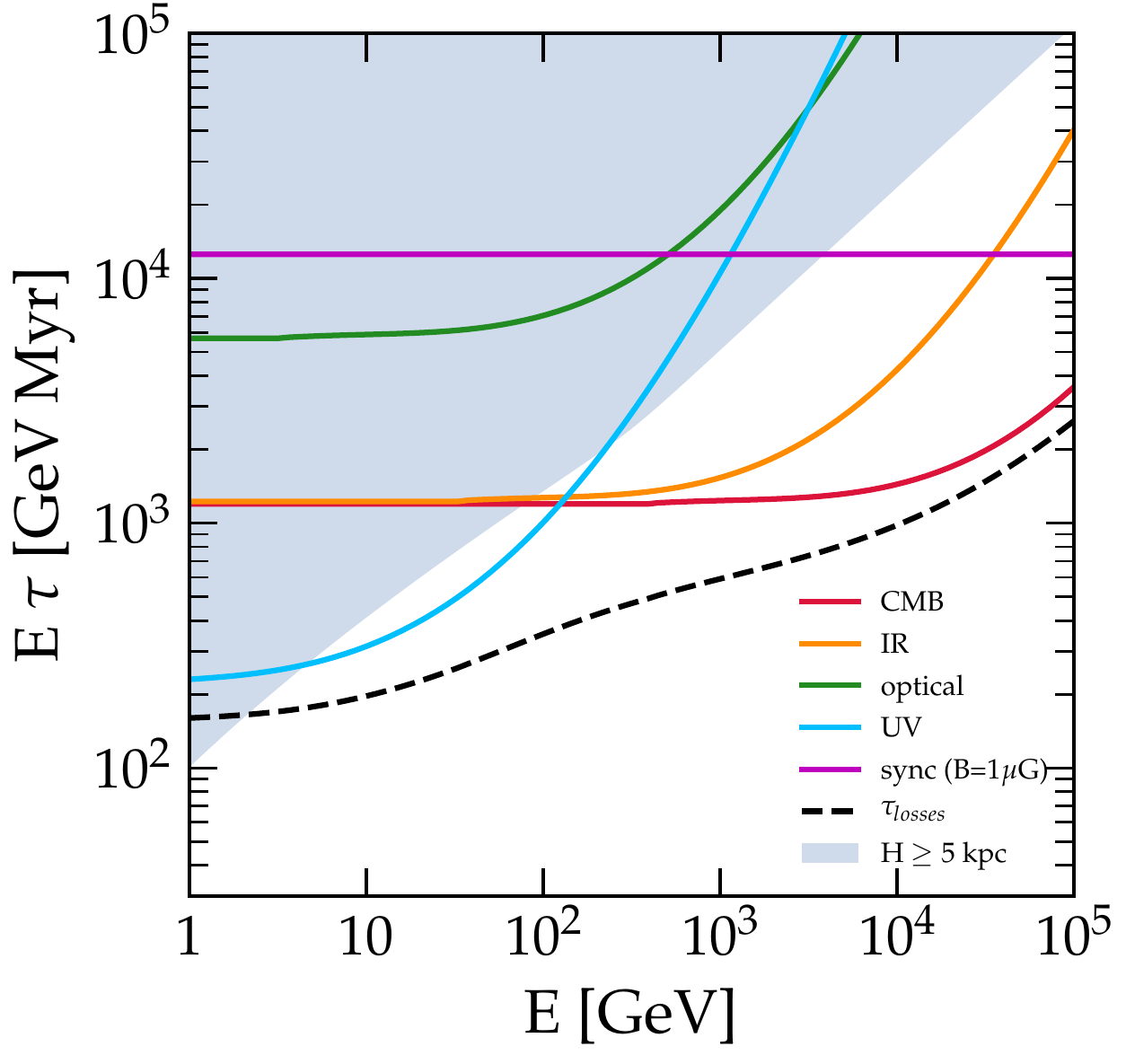}
\hspace{\stretch{1}}
\includegraphics[width=0.48\textwidth]{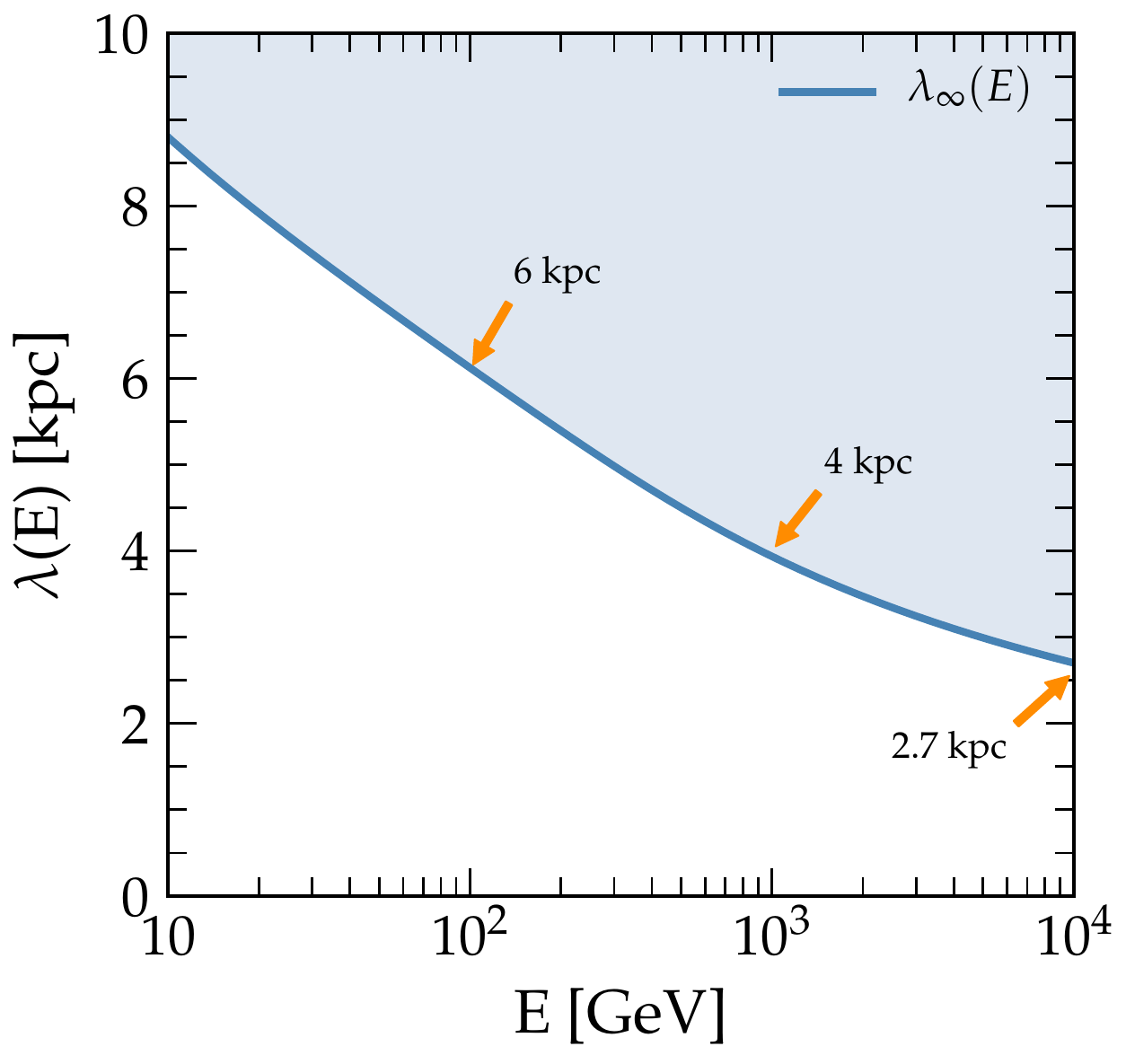}
\hspace{\stretch{1}}
\caption{Left panel: energy loss timescale for CR electrons as a function of energy during their propagation in the Galaxy. The timescales are multiplied by $E$ to give prominence to the deviations from the standard $b \propto E^2$ regime. The dashed line represents the total loss time, while the solid lines refer to the individual contributions of the magnetic field (magenta line) or ISRF components. The shadowed region marks out the escape timescale from the Galaxy due to diffusion. Right panel: the average mean free path in the Galaxy of a lepton with energy $E$.}
\label{fig:losses}
\end{figure*}

The transport of CR electrons and positrons in the Galaxy is governed by diffusion, convection, second-order Fermi acceleration and energy losses. If we restrict ourselves to energies $\gtrsim 20$ GeV, the role of advection and reacceleration (if any) can be neglected, and the transport equation reduces to the simpler form~\cite{TheBible}:
\begin{equation}\label{eq:transport}
\frac{\partial}{\partial t} n_e(t, E, \vec r) = D(E) \nabla^2 n_e(t, E, \vec r) - \frac{\partial}{\partial E}\left[ b(E) n_e(t, E, \vec r) \right] + {\cal Q}(t, E, \vec r) \, .
\end{equation}
Here, $n_e(\vec r,t,E) = dN/dVdE$ is the isotropic part of the differential CR lepton density and is related to the differential flux as $\Phi = (d^4N)/(dEdAdtd\Omega) = n_e c/ 4\pi$ since we used the assumption that the particles we are dealing with are relativistic.
The source density rate is ${\cal Q}(\vec r,t,E) \equiv dN/dE dV dt$, $D(E)$ is the energy-dependent isotropic diffusion coefficient, and $b(E) \equiv dE/dt$ is the rate of energy losses. 
For simplicity we assume that both diffusion and energy losses are spatially uniform through the Galaxy.
As usual, Eq.~\ref{eq:transport} is solved with the so-called free-escape boundary condition at $z=|H|$, $H$ being the height of the halo, namely $n_e(z = \pm H) = 0$. We neglect particle escape in the radial direction, a good approximation so far as the distance of the Sun to the radial boundary is larger than the size $H$ of the halo.
The diffusion-advection equation can also be written for nuclei and the transport parameters, most notably the diffusion coefficient and the halo size, can be tuned to the observed secondary-to-primary flux ratios~\cite{Evoli2019prd,Weinrich2020aab}. 
Following~\cite{Evoli2019prd}, we adopt for the diffusion coefficient a broken power-law functional form:  
\begin{equation}
D(E) = 2 v_A H + D_0 \left(\frac{E}{\rm GeV}\right)^\delta 
\left[1 + \left(\frac{E}{E_0} \right)^{\Delta \delta / s}\right]^{-s},
\end{equation}
where $s = 0.1$, $\Delta \delta = 0.2$, and $E_0 = 312$~GeV are, respectively, the smoothing, the magnitude and the characteristic energy of the break, and we adopt for these parameters their best-fit values as derived from fitting the H and He spectra. The slope of the diffusion coefficient $\delta$ and the Alfv\'en velocity $v_A$ can be fitted to the ratios of secondary to primary nuclei, mostly boron to carbon (B/C) and boron to oxygen (B/O). As far as the normalization $D_0$ and the halo size $H$ are concerned, only their ratio can be constrained by this kind of analysis, rather than the value of each independently. On the other hand, using the flux of unstable isotopes (in particular $^{10}$Be), the halo size $H$ can be effectively constrained, although with much larger uncertainty~(see~\cite{Evoli2020prd,Weinrich2020aaa}).
In the following, we use the parameter values provided by~\cite{Evoli2020prd}, where the halo size was found to be $H \gtrsim 5$~kpc. We fix $H = 5$~kpc for the sake of definitiveness, and then $D_0 = 2.2 \times 10^{28}$~cm$^2$/s, $\delta = 0.54$, and $v_A = 5$~km/s.

The transport of leptons on Galactic scales is dominated by the role of energy losses, mainly inverse Compton scattering (ICS) on the photons of the interstellar radiation fields (ISRFs) and cosmic microwave background (CMB), and synchrotron emission in the Galactic magnetic field~\cite{Delahaye2010aa}. 
For the magnetic field we assume U$_{\rm B} = 0.025$~eV~cm$^{-3}$ (corresponding to a magnetic field B$_{0} = 1$~$\mu$G), as representative of the average energy density in the halo~\cite{Delahaye2010aa}. 
The ISRF is made of the CMB, whose energy density is U$_{\rm CMB} = 0.25$~eV~cm$^{-3}$ everywhere, and of a second component which is the result of emission by stars and reprocessing of the starlight by dust in the ISM. In order to obtain accurate models for the photon distribution and spectrum of this component, it is therefore necessary to perform a detailed modeling of the Galactic stellar populations distribution, dust distribution, and the absorption and scattering of light.
In an approach based on the Green function formalism, the ISRF spatial structure cannot be taken into account, and one has to rely on some spatial average of the ISRF on scales of the order of the propagation mean free path for electrons of energies above 10 GeV.
The space-dependent ISRF model provided by the authors of the GALPROP code~\cite{Moskalenko2006apj} was averaged on a cylinder of radius and half-height of 2 kpc around the Sun by~\cite{Delahaye2010aa}, showing that a reasonable fit of the averaged ISRF as a function of the photon energy is given by the sum of five gray bodies which are identified with an infrared (IR), an optical and three ultraviolet components (UV-I, UV-II, and UV-III).
These authors also discussed the anticorrelation between the IR and UV components in terms of spatial distributions. In particular, considering a smaller volume around the Sun results in a larger (smaller) IR (UV) contribution, due to the efficient UV-absorption and IR-emission properties of the dust, which is mostly concentrated in the disk~\cite{Delahaye2010aa}.
As we will show afterwards, the mean free path of $\sim 10-100$ GeV electrons is larger than 2 kpc, as it becomes comparable with the halo size $H$ around 10 GeV. Therefore, in our calculation it is reasonable to average the UV and IR light on larger regions than done in previous work. As a consequence, the relative contribution of the UV energy density with respect to the IR is expected to be larger. On the other hand, at larger distances from the Sun, the actual distribution of stars and dust is more uncertain and different assumptions in modeling these contributions can result in wide differences in the UV energy density. To quantify the UV-to-IR energy density ratio on a distance of the order of the halo size, $H\gtrsim 5$~kpc, we exploited the publicly available ISRF distribution provided by~\cite{Vernetto2016prd} and found that on a scale of $\sim 5$~kpc around the Sun the UV-to-IR ratio is estimated to be a factor $\sim 2$ higher than the one derived by~\cite{Delahaye2010aa} in a smaller box. In the calculations below, we adopt the result of such an averaging procedure for the UV and IR backgrounds, which implies a UV background roughly twice as large as the one used by~\cite{Delahaye2010aa}.
  
The rate of losses can be written as
\begin{equation}\label{eq:bE}
b_{\rm e}(E) = \frac{4}{3} c \sigma_T \left[ \sum_i f_{\rm KN}(E, T_i) U_{\gamma,i} + U_B \right] \left(\frac{E}{m_e c^2}\right)^2  
\end{equation}
where $m_e$ is the electron mass, $\sigma_T$ is the Thompson scattering cross section and $U_{\gamma,i}$ and $U_B$ are the energy densities in the photons of type $i=\{\text{ISRF},~\text{CMB}\}$ and in the form of magnetic field respectively. 
The function $f_{\rm KN}$ effectively describes the modification to the Thomson cross section due to the KN corrections and we adopt the recent parametrization given in~\cite{Fang2020arxiv}.

In Fig.~\ref{fig:losses} (left panel) we show the timescale for energy losses $\tau_{\rm loss} = E / b(E)$ as a function of the lepton energy, compared with the timescale for diffusive escape from the Galaxy $\tau_{\rm esc} = H^2 / 2D(E)$. The shaded area shows the escape time for halo size $H$ larger than $5$ kpc, as derived from the analysis of unstable elements~\cite{Evoli2020prd}. 
The loss time scale as due to ICS off the optical (green line), infrared (orange line), UV (blue line) and CMB (red line), and the one associated with synchrotron emission (purple line) are shown separately, while the black dashed line shows the total loss time. Several comments are worth making: (i) in the energy region of interest here, ICS enters the KN regime only for scattering off optical light (at $\sim 2$ TeV) and UV light (at $\sim 50$ GeV); (ii) the total loss time is shorter than the diffusion timescale for all energies of interest here; (iii) the total loss time depends on energy in a nontrivial way and shows a feature in the neighborhood of the transition of ICS to the KN regime on the UV background. As discussed in detail in Ref.~\cite{Evoli2020prl}, this reflects into a corresponding feature of the electron spectrum as observed by AMS-02.

It is worth noticing here that in the calculations that we discussed in Ref.~\cite{Evoli2020prl}, we adopted a parametrization of the transition of ICS to the KN regime that was recently criticized by~\cite{Fang2020arxiv}. There are several such parameterizations (see for instance~\cite{Schlickeiser2010njph,Aharonian1985afz,Hooper2017prd,Stawarz2010apj}) that describe the transition to different levels of accuracy and using different assumptions and different functional forms. In Ref.~\cite{Evoli2020prl} we used the one put forward by~\cite{Hooper2017prd} (numerically similar to that of Ref.~\cite{Schlickeiser2010njph}), which turns out to describe the transition rather poorly, as correctly pointed out by~\cite{Fang2020arxiv}, making it sharper than the exact solution shows. Nevertheless, this is more a quantitative issue than a qualitative one: as we show below, even using the parametrization proposed by~\cite{Fang2020arxiv}, the feature observed in the CR electron data is still present in the calculated electron spectrum, although a somewhat different and apparently better justified choice of the ISRF is required (see discussion above).

The energy losses suffered by electrons imply that for a given energy $E$ only particles located within a given distance can contribute to the flux at Earth. This distance is the one covered by an electron under the effect of losses and diffusion, $\lambda_{\infty}(E)$:
\begin{equation}
\lambda_\infty^2(E) \equiv 4 {\int_E^{\infty} dE' \, \frac{D(E')}{|b(E')|}},
\end{equation}
obtained assuming that the source is able to provide particles with arbitrarily high energy. Hence $\lambda_{\infty}(E)$ provides an upper limit to the distance of a source of electrons of energy $E$. The propagation horizon for CR leptons in the Galaxy, $\lambda_\infty$,  is shown in Fig.~\ref{fig:losses} (right panel) as a function of the CR observed energy $E$. The shaded area is obtained by allowing for halo size $H>5$ kpc but rescaling the diffusion coefficient in such a way that $D/H$ remains unchanged, so as to retain the correct B/C and B/O ratios.

If the sources are assumed to occur at a rate $\mathcal R$ and have a short duration compared with the typical timescales for diffusion and losses (bursting sources), the assumption that they are homogeneously distributed in space is reasonably good as long as the number of sources that contribute at a given energy is $\gg 1$. On the other hand, when $\mathcal R \lambda_\infty^2(E) \tau_{\rm loss}(E) / R_d^2 \sim$ a few, the assumption breaks down and the flux of leptons observed at any given location becomes dominated by fluctuations associated with the most recent and closest sources. Given the high precision of data that became available with AMS-02, these fluctuations need to be taken into account. The only way to do so, at present, is to solve the transport equation using the Green function formalism, so that the flux observed at Earth can be split in the contribution provided by each source at a given time. For pointlike sources located at a position $\vec r_s = (x_s, y_s, z_s)$ and injecting leptons with energy $E_s$ at a time $t_s$, the Green function of Eq.~\ref{eq:transport} can be written as follows~\cite{Syrovatskii1959}:		
\begin{equation}
{\mathcal G} (t, E, \vec r \leftarrow t_s, E_s, \vec r_s) = \delta(\Delta t - \Delta \tau) {\mathcal G}_{\vec r} (E, \vec r \leftarrow E_s, \vec r_s)
\end{equation}
where the $\delta$ function shows that a particle injected with energy $E_s$ is observed after a time $\Delta t \equiv t - ts$ with energy $E < E_s$ only if the elapsed time corresponds to the time during which the energy of a particle decreases from $E_s$ to $E$ because of losses. This loss time is defined as 
\begin{equation}
\Delta \tau(E, E_s) \equiv \int_E^{E_s} \frac{dE'}{b(E')} \,.
\end{equation}
The spatial part of the Green function that satisfies the correct boundary conditions at $z = \pm H$ can be obtained by using the image charge method~\cite{Cowsik1979apj}:
\begin{equation}
{\mathcal G}_{\vec r} = \frac{1}{b(E)} 
\frac{1}{\pi \lambda_e^2} \exp\left[ -\frac{(x-x_{s})^{2}+(y-y_{s})^{2}}{\lambda_e^2}\right] 
\frac{1}{\sqrt{\pi}\lambda_e} \sum_{n=-\infty}^{+\infty} (-1)^{n} \exp \left[ -\frac{(z-z^{(n)}_{s})^{2}}{\lambda_e^2} \right],
\end{equation}
where $z^{(n)}_s = 2 n H + (-1)^n z_s$. We checked that the infinite series can be truncated at $n_{\rm max} = 10$ retaining an accuracy better than $10^{-6}$.

In the following we use the notation:
\begin{eqnarray}
\Sigma(x,y,\lambda_e) & \equiv & \exp\left[ -\frac{(x-x_{s})^{2}+(y-y_{s})^{2}}{\lambda_e^2}\right] \\
\chi(z,\lambda_e) & \equiv & \sum_{n=-\infty}^{+\infty} (-1)^{n} \exp \left[ -\frac{(z-z^{(n)}_{s})^{2}}{\lambda_e^2} \right]
\end{eqnarray}

\begin{figure*}[t]
\hspace{\stretch{1}}
\includegraphics[width=0.47\textwidth]{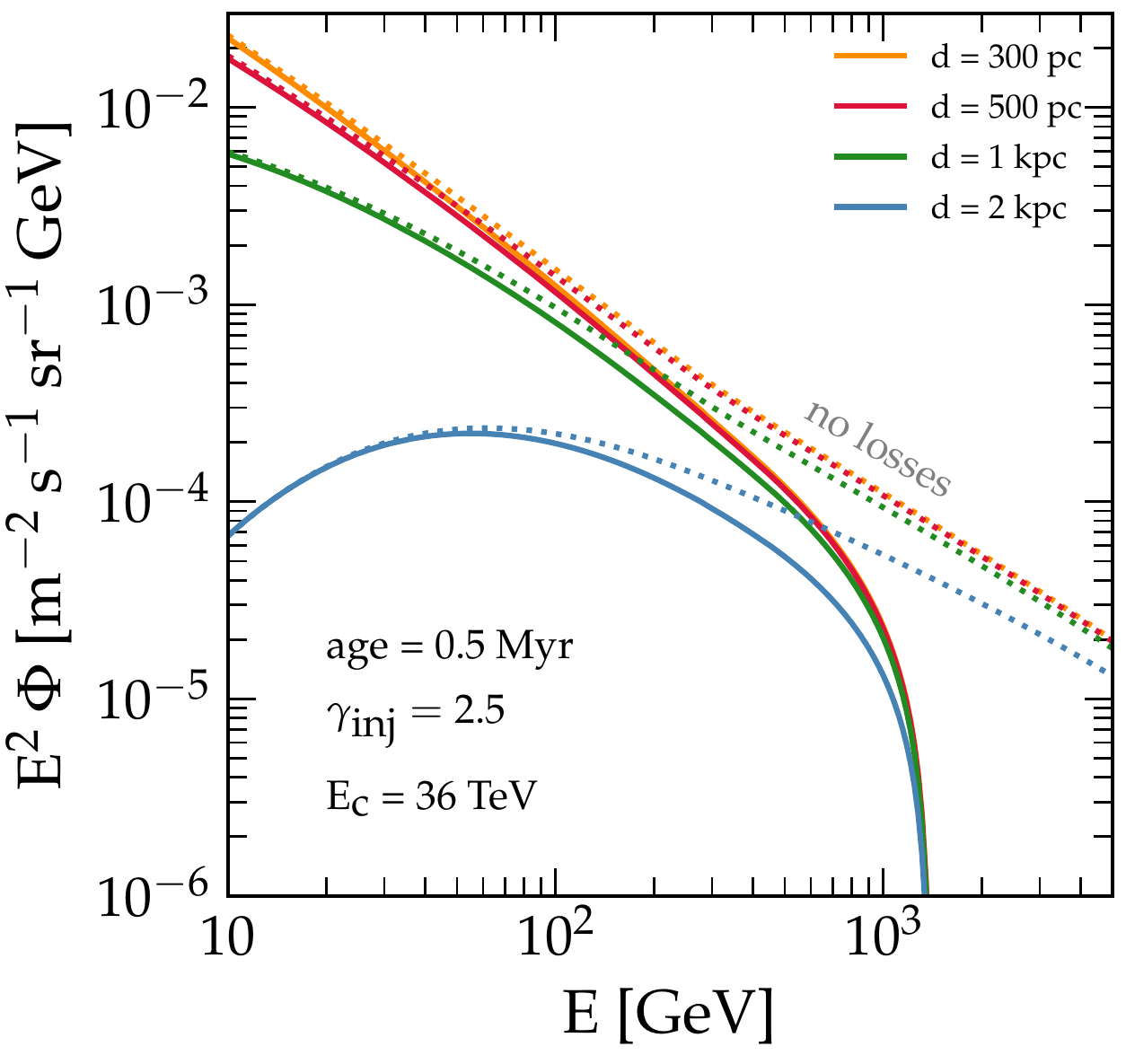}
\hspace{\stretch{1}}
\includegraphics[width=0.47\textwidth]{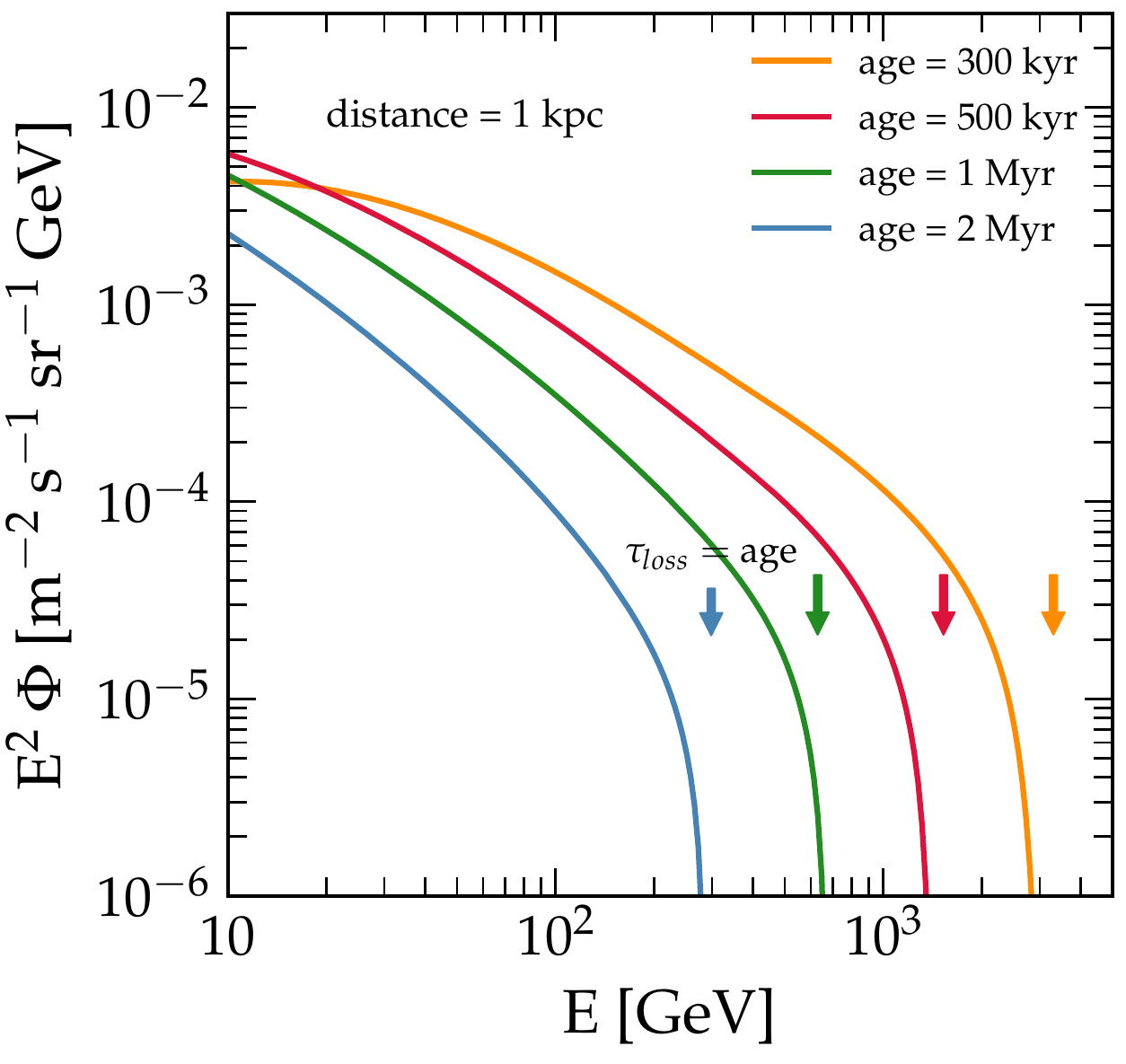}
\hspace{\stretch{1}}
\caption{Left panel: electron fluxes computed from a single SNR of given age, $t = 500$~kyr, assumed to be located at different distances from $d = 0.3$ to $d = 2$~kpc as in legend. The dotted lines show the corresponding solution of the transport equation in the absence of energy losses. Right panel: electrons from a single SNR at the given distance $d = 1$~kpc and different ages from $t = 0.3$ to $t = 2$~Myr. The arrows mark the energy at which the loss timescale $\tau_{\rm loss}$ equals the source age.}
\label{fig:singleSNR}
\end{figure*}

Finally, the generic solution of Eq.~\ref{eq:transport} evaluated at the Sun location, which we fix to be $(x_\odot, y_\odot, z_\odot) = (8.5, 0, 0)$~kpc, and at the time $t_\odot$ corresponding to the total time of the simulation, is obtained in terms of the Green function by integrating over a distribution in energy, space and time of sources as  
\begin{equation}\label{eq:formalsolution}
n(t_\odot, E, \vec r_\odot) = \int \!\!\! \int \!\!\! \int dt_s \, dE_s \, d^3 \vec r_s \, \delta(\Delta t - \Delta \tau)
{\mathcal G}_{\vec r}(E, \vec r_\odot \leftarrow E_s, \vec r_s) {\cal Q} (t_s, E_s, \vec r_s).
\end{equation}

In the realistic calculations discussed below the positions of the sources are generated at random in such a way as to follow the spiral structure of the Galaxy and with a rate that reflects the rate of occurrence of the type of sources we are interested in. Notice also that Eq.~\ref{eq:formalsolution} allows us to introduce sources that are not necessarily burstlike, namely have a finite duration in time. This is especially useful in describing the behavior of pulsar wind nebulae as lepton sources, since their typical spin-down timescales can be rather long. 

\section{Galactic Sources of leptons}
\label{sec:leptonsources}

High-energy electrons and positrons are produced in the Galaxy in at least three different ways: (i) Electrons are accelerated at SNR shocks, together with their hadronic counterparts. This is confirmed by the observation of IC-scattered gamma rays, radio and nonthermal x-ray emission from SNRs, and well accounted for by state-of-the-art numerical simulations of diffusive shock acceleration.
The latter could make sense of the observed flux ratio between hadrons and electrons in the GeV region, of order $\sim 100$, in terms of different injection mechanisms of the two species \cite{Jaehong2015prl}. In general, because of the radiative losses suffered by electrons inside the parent remnant, the spectrum of electrons released into the ISM can be different from that of protons~\cite{Diesing2019prl}. 
(ii) Electrons and positrons are produced in about the same amount in pulsar magnetospheres and eventually accelerated at the pulsar wind termination shock (see \cite{Amato2020} and references therein). Pulsars have typically a longer lifetime compared with SNRs and guarantee an approximately equal number of electrons and positrons with a relatively hard spectrum up to a few hundred GeV. 
(iii) Finally, electrons and positrons are generated as secondary products of inelastic hadronic interactions during CR transport in the Galaxy. The spectrum of these particles is straightforward to calculate once the parent proton spectrum and the diffusion coefficient are given. 

In principle one should also take into account the contribution of electrons and positrons produced as secondary products of CR interactions inside SNRs~\cite{Berezhko2003aa,Blasi2009prl}. We assume here that the average grammage traversed during the lifetime of a SNR is small compared with the one associated to Galactic CR transport. In fact, we now know that if this were not the case, we should have observed the contribution of this process to the antiproton \cite{BlasiSerpico2009prl} and secondary nuclei spectra~\cite{Mertsch2009prl} in the cosmic radiation, which does not seem to be the case, based on AMS-02 observations~\cite{AMS02.2016.pbar,AMS02.2018.libeb}. 

In this section we discuss our calculations of the three source terms for leptons in the Galaxy, electrons from SNRs (Sec.~\ref{sec:primarye}), electrons and positrons from pulsars (Sec.~\ref{sec:pulsarmodel}) and secondary electrons and positrons (Sec.~\ref{sec:secondaryep}).

\subsection{Primary electrons from SNRs}
\label{sec:primarye}

The escape of CRs from a SNR is a very difficult problem to tackle and clearly affects the spectrum of the particles (both protons and electrons) liberated into the ISM. Recent work~\cite{Cristofari2020aph} has shown how the spectrum of protons as due to escape from the upstream and release of the bulk of protons at the end of Sedov phase, after adiabatic losses, may be rather complex and reflect the inhomogeneity of the environment in which the explosion takes place. 
For electrons, the maximum energy is typically determined by energy losses; hence, the escape from upstream is expected to be less of an issue than for protons. On the other hand, radiative losses during the supernova expansion may lead to a different injection spectrum of protons and electrons~\cite{Diesing2019prl}. 
Here, for simplicity, we assume that, in terms of electron injection into the ISM, each SNR acts as an impulsive source with a given spectrum. This assumption is justified since the lifetime of a SNR as an accelerator is typically $10-20$ thousand years, much shorter than the diffusion time of electrons from the closest sources.

The injection term for a SNR event located in the Galactic disk at a position $\vec r_0$ and at time $t_{0}$ is modeled with a source term as 
\begin{equation}\label{eq:qburst} 
{\mathcal Q}(t_s, E_s, \vec r_s) = \delta^3(\vec r_s - \vec r_0) \delta(t_s - t_{0}) Q_{\rm SNR}(E_s).
\end{equation}
The energy-dependent part is chosen so as to reflect the spectrum of electrons accelerated at a SNR in the presence of energy losses. 
As discussed in Refs.~\cite{Zirakashvili2007aa,Blasi2010mnras} the shape of the cutoff depends on the diffusion coefficient in the acceleration region. 
For Bohm diffusion, most reasonable in the case of strong magnetic field amplification, the spectrum reads:
\begin{equation}
Q_{\rm SNR}(E) = Q_{0} \left(\frac{E}{\rm GeV}\right)^{-\gamma} \exp\left[-\left(\frac{E}{E_c}\right)^2\right],
\end{equation}
where the normalization $Q_0$ and the injection slope of this primary component, $\gamma \gtrsim 2$, is chosen in such a way as to reproduce local observations.
We define the acceleration efficiency as the fraction $\xi$ of the kinetic energy of the supernova ejecta, $E_{\rm SN} = 10^{51}$~erg, that is converted to relativistic electrons with $E>100$ MeV (arbitrary normalization point) through diffusive shock acceleration: 
\begin{equation}
\xi E_{\rm SN} = \int_{100 \, \rm MeV}^\infty dE \, E Q(E).
\end{equation}

The cutoff energy $E_c$ is set by equating acceleration and losses timescales in the acceleration region:
\begin{equation*}
t_{\rm acc} \sim t_{\rm loss} \rightarrow \frac{8D_B}{u_s^2} = \frac{E}{\dot{E}}.
\end{equation*}

For Bohm diffusion ($D_{\textrm B}$) and synchrotron losses in a magnetic field of $\sim$0.1~mG, typical conditions for the environment downstream of a SNR shock~\cite{Vink2012aarv}, the electron spectrum develops a cutoff at $E_{\rm c} \simeq 36$~TeV. However, this is not a critical parameter for our calculations, since energy losses can produce a cutoff in the propagated spectrum at much lower energy.

By inserting the source term in Eq.~\ref{eq:qburst} in the formal solution of the transport equation (Eq.~\ref{eq:formalsolution}), we obtain that the contribution of the $i$-th source to the local electron density is different from $0$ only if the time since the electron release $t_\odot - t_{0}$ is shorter than the loss timescale $\tau_{\rm l}(E) \equiv \Delta \tau(E, \infty)$. In this case this contribution is given by 
\begin{equation}\label{eq:burstsolution}
n_i^{\rm SNR}(t_\odot, E, \vec r_\odot) = \frac{Q_{\rm SNR}(E_*)}{(\pi \lambda^2_*)^{3/2}} \frac{b(E_*)}{b(E)} \Sigma(x,y,\lambda_*) \chi(z,\lambda_*)
\end{equation}
where $E_*$ is obtained by inverting $t_\odot - t_{0} - \Delta \tau(E, E_*) = 0$ and $\lambda_* = \lambda_e(E, E_*)$.

In Fig.~\ref{fig:singleSNR} we show the contribution of a single source to the local flux, fixing its age and changing its distance $d \equiv \| \vec r_\odot - \vec r_0 \|$ (left panel) and fixing the distance and changing the source age (right panel). In these plots we assume an injection slope $\gamma = 2.5$ and an acceleration efficiency $\xi = 0.1$\%.
For bursting sources, the maximum energy that particles can reach Earth with is the one for which the age of the source $t_a$ is approximately equal to the loss time. This is illustrated very clearly in the left panel of Fig.~\ref{fig:singleSNR}, where the source distance is changed while the age is fixed, and all spectra exhibit a cut off at the energy where $\tau_{\rm loss}\approx t_{a}$. Figure~\ref{fig:singleSNR} also illustrates a few other interesting points: the cases with and without energy losses are shown and one can clearly appreciate how in the latter case the flux at Earth scales as $Q(E)/(D(E) t_{a})^{3/2}$. The slight curvature in the corresponding curves reflects the change of slope in the diffusion coefficient adopted for CR transport in the Galaxy. On the other hand, for energies that are low enough that the diffusion time from the source to the Earth is comparable or longer than the age of the source, particles cannot reach the Earth, so that a low-energy flux drop is expected. This drop occurs at higher energies for more distant sources. The effect is clearly visible in both panels of Fig.~\ref{fig:singleSNR} and occurs in much the same way whether there are losses or not. In the most general case the position of the drop, $E_{\rm diff}$, can be found by solving the equation $\lambda(E_{\rm diff}, E_*) = d$. 

It is worth asking what are the conditions that are necessary for an astrophysical source to be approximated as a burst. Since the arrival of particles even from an impulsive source is spread over a time comparable with the diffusion timescales from the source to a distance $d$, the condition that needs to be fulfilled is that the duration of the source is much smaller than the diffusive time. Clearly this criterion is most constraining at high energies where the diffusion time is the shortest. For all practical purposes, a SNR with an age of a few $\times 10^4$ yr can be considered as a burst. 
As far as the spatial extent of the sources is concerned, the typical size of an active SNR is of order $\sim$~10 pc which is significantly smaller than all the other scale-lengths of the problem (e.g., the mean free path of the particles we are dealing with, the average distance from sources, etc.) therefore we can safely ignore their spatial extent and assume they are point-like.
	
\subsection{Electrons and positrons from pulsars}
\label{sec:pulsarmodel}

Pulsars are known to be powerful sources of $e^\pm$ pairs (see~\cite{Bykov2017ssrv} for a recent review).
The pairs are produced in the neutron star magnetosphere, where the electrons originally extracted from the surface fly away from the star along curved field lines. The photons these primary electrons emit, either by curvature radiation or by ICS on the star radiation field, are energetic enough for pair production in the local magnetic field. The process results in a cascade, that for typical pulsar parameters leads to the production of $\sim 10^{4}-10^{6}$ pairs for each electron: this is the so-called multiplicity factor. These pairs leave the star vicinities in the form of a relativistically expanding magnetized wind, which needs to slow down in order to match the nonrelativistic motion of the confining medium, either the SN ejecta, as in the case of young pulsars still residing within the associated SNR, or the ISM, as in the case of more evolved systems, in which the pulsar has already left the SNR and is moving supersonically through the ISM, producing a bow shock. The dissipation of the wind bulk motion happens at a termination shock, where particles are accelerated with extraordinary efficiency. While the acceleration process is not clear (see e.g. \cite{Amato2020} for a review of the proposed mechanisms), observations of these systems typically point to a pair population with a broken power-law spectrum, with index $\gamma_L<2$ for energies $E<E_b \sim 500$ GeV and $\gamma_H>2$ above $E_b$ (see e.g.~\cite{Bucciantini2011mnras}).

Pulsars were proposed as sources of positrons long before~\cite{Harding1987icrc,Atoyan1995prd} the PAMELA discovery of the positron excess \cite{PAMELA.2009.posfraction}. After such discovery it was soon recognized that data could be described very naturally in terms of the positron contribution from old neutron stars \cite{Hooper2009jcap}, but the properties required for pulsars to be good candidate sources and the issue of the escape of the pairs from the magnetosphere remained open. The crucial role of middle-aged neutron stars with large kick velocities, able to escape the parent remnant within a few thousand years of the explosion was first discussed in \cite{BlasiAmato2011}: these pulsars would generate a BSN in the ISM and the pairs may be able to leave the magnetosphere from the tail of the nebula. In the same article the authors described the spectrum of the low-energy electrons, most important for the explanation of the positron fraction observed by PAMELA~\cite{PAMELA.2009.posfraction} and more recently by AMS-02~\cite{AMS02.2013.posfraction}.

The emission of $e^\pm$ pairs from pulsars has recently been confirmed by the detection by the HAWC observatory of diffuse gamma-ray emission, apparently due to inverse Compton scattering of extremely high-energy leptons from extended regions around selected pulsar wind nebulae~\cite{HAWC.2017.tevhalos}. These observations, in addition to supporting the picture in which high energy leptons may leak out of pulsar wind nebulae, also provided evidence for slow diffusion in the circumnebular region, probably associated to enhanced turbulence level in the same region. In fact the diffusion coefficient in these regions, typically of size $\lesssim 100$ pc around the source, is about 2 orders of magnitude smaller than estimated on Galactic scales. This enhanced turbulence level is very important for explaining the diffuse gamma-ray emission from the vicinities of these sources; it remains to be seen whether they can modify the spectra of electrons and positrons with respect to the source spectra. However it seems plausible that a cutoff in the TeV energy range may be produced due to radiative losses in these regions. Since we do not know yet how general the cocoons of enhanced turbulence are, we do not include them in an explicit manner. Nevertheless the source spectra that we will find as a result of our calculations should be considered as the spectra of particles leaving such regions, if they are generic enough.

Here we model the emission of $e^\pm$ pairs from pulsars by adopting a minimal set of assumptions and making use of the few pieces of information that are well rooted in observations. When this is not possible, we justify the assumptions made. 

First we assume that the pulsar spin-down energy is dissipated via magnetic dipole radiation (braking index of $n = 3$). This assumption is in fact not well justified but difficult to avoid. Values of the braking index very different from 3 have been observed in several objects and are possibly explained as a result of variations of the pulsar internal properties (see \cite{Parthasarathy2020mnras} and references therein). However, models of the pulsar population (as, e.g., the one by \cite{FaucherGiguere2006apj}, which we will use in the following) are based on this assumption and without it several quantities become ill defined.

For magnetic dipole spin down, the upper limit to the luminosity in the form of pairs reads:
\begin{equation}\label{eq:luminosity_t}
\mathcal{L}_{e^\pm}(t) = \frac{1}{2} I \Omega_0^2 \frac{1}{\tau_0} \frac{1}{\left( 1 + \frac{t}{\tau_0} \right)^2},
\end{equation}
where $I = \frac{2}{5} M_{\rm S} R_{\rm S}^2$ and $\tau_0 = \frac{3 c^3 I}{B_{\rm S}^2 R_{\rm S}^6 \Omega_0^2}$, are given in terms of $M_{\rm S} = 1.4 M_\odot$, $R_{\rm S} = 10$~km and $B_{\rm S} = 10^{12.65}$~G~\cite{FaucherGiguere2006apj}.
Here $P_0$ is the initial rotation period of the pulsar and $\Omega_0 = \frac{2\pi}{P_0}$ is the corresponding angular frequency.

Numerically, the spin-down age $\tau_{0}$ can be written as  
\begin{equation}
\tau_0 \sim 36 \left( \frac{P_0}{0.1 \, {\rm s}} \right)^2 \, {\rm kyr}
\end{equation}

The source term for positrons and electrons from an individual pulsar, assuming continuous injection as a function of time, is:
\begin{equation}\label{eq:source-pwn}
{\cal Q}(t_s, E_s, \vec r_s) = \delta^3(\vec r_s - \vec r_0) Q_{\rm PWN}(E_s, t_s) 
\end{equation}
where, following \cite{AmatoBlasi2018}, the spectrum $Q_{\rm PWN}(E)$ is modeled as a broken power law, with slope $\gamma_{\rm L}$ below the break $E_b$ and slope $\gamma_{\rm H}$ above:
\begin{equation}
Q_{\rm PWN}(E, t) = Q_0 (t) 
{\rm e}^{-E/E_{\rm c}(t)}
\times
\begin{cases}
(E / E_{\rm b})^{-\gamma_{\rm L}} & E < E_{\rm b} \\
(E / E_{\rm b})^{-\gamma_{\rm H}} & E \ge E_{\rm b}.
\label{eq:injpulsar}
\end{cases} 
\end{equation}

In most cases, observations of electromagnetic radiation from individual pulsars requires $\gamma_{\rm L} \sim 1-1.9$ and $\gamma_{\rm H}\sim 2.5$. This functional form provides a good description of the emission from PWNe both within the parent SNR and in the bow shock phase \cite{Bykov2017ssrv}. 
The cutoff position $E_{\rm c}$ is assumed to be a function of time reflecting the temporal evolution of the potential drop $V$~\cite{Kotera2015jcap}:
\begin{equation}\label{eq:cutoff_t}
V(t) = \frac{2 \pi^2 B_S R_S^3}{c^2 P_0^2} \frac{1}{1 + t / \tau_0}. 
\end{equation}
Numerically, the cutoff energy reads 
\begin{equation}
E_{\rm c}(t) \sim 3 \, \text{PeV} \, \left( \frac{P_0}{0.1 \, {\rm s}} \right)^{-2} \frac{1}{1 + t / \tau_0}. 
\end{equation}

\begin{figure}
\centering
\includegraphics[width=0.48\textwidth]{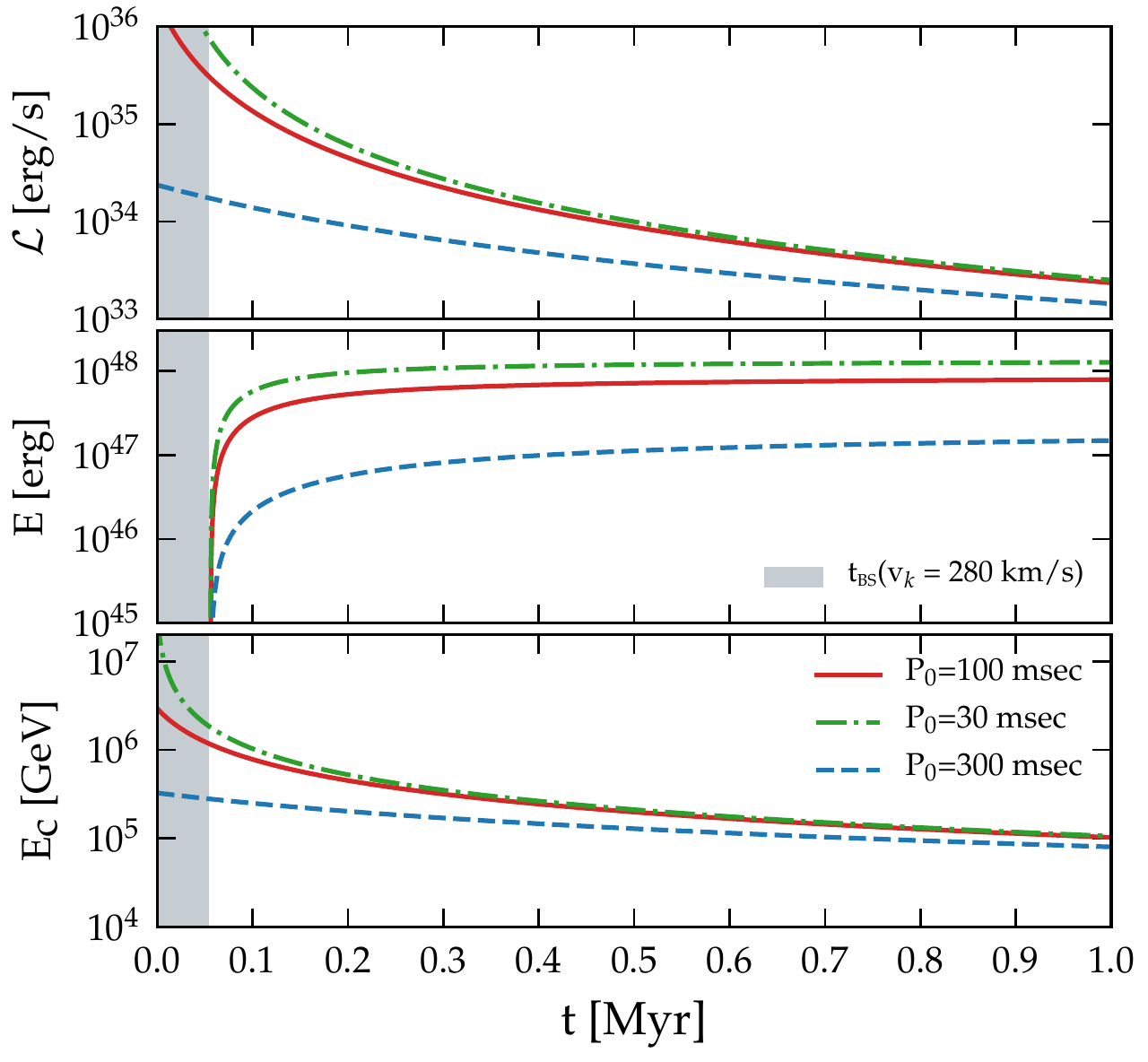}
\caption{The spin-down luminosity (top panel), total energy released (middle panel), and cutoff energy (bottom panel) as a function of time for a pulsar with $P_0 = 30$~ms (dash-dotted green curves), $P_0 = 100$~ms (solid red curves), and $P_0 = 300$~ms (dashed blue curves). We also assume $v_k = 280$~km/s and $\epsilon = 1$. The shadow area corresponds to the $t_{\rm BS}$ when the pulsar is still inside the SNR.}
\label{fig:pulsar}
\end{figure}

The spectrum normalization $Q_0(t)$ is obtained at any given time by assuming that a fraction $\epsilon$ of the spin-down luminosity is converted to pairs:
\begin{equation}
\epsilon \mathcal{L}_{e^\pm}(t) = \int_{0}^\infty \!\! dE \, E Q(E),
\end{equation}
and is distributed in equal parts between electrons and positrons. Notice that since $\gamma_{\rm L}<2$ and $\gamma_{\rm H}>2$ the energetics of pairs is dominated by the region around $E_{\rm b}\sim 100-1000$~GeV.

We assume continuous injection for $t_s > t_0 + t_{\rm BS}$, where $t_{\rm BS}$ is the time when the pulsar leaves the SNR due to its proper motion and eventually forms a BSN. As mentioned above, in this phase, the termination shock is formed due to the interaction between the relativistic pair wind and the ISM in which the pulsar moves. 
The pulsar leaves the parent SNR at a time such that $v_k t=R_{\text{sh}}(t)$, where $v_{k}$ is the birth kick velocity and $R_{\text{sh}}$ is the position of the forward shock associated with the SN explosion, as a function of time. The latter depends on the environment in which the explosion takes place and may have a rather complex time dependence. Here we consider the easiest estimate of the escape time of the pulsar from the remnant by assuming that the explosion takes place in a constant density ISM with density $n_0 = 3$~cm$^{-3}$. We also assume an explosion energy of $10^{51}$~erg. Typically the escape of the pulsar from the remnant takes place during the Sedov-Taylor phase \cite{Taylor1950,Sedov1959}, when the radius of the shell increases with time as $R_{\text{sh}}(t) \simeq 3.2 (E_{51} / n_0)^{1/5} \, (t / \text{kyr})^{2/5}$~pc. Hence a pulsar with typical birth velocity $v_k = 280$~km/s crosses the forward shock at a time:
\begin{equation}
t_{\rm BS} \simeq 56 \, \text{kyr} \, \left( \frac{E_{51}}{n_0} \right)^{1/3} \left(\frac{v_k}{280 \, \text{km/s}}\right)^{-5/3}.
\end{equation} 

The total energy in the form of pairs released after the pulsar has crossed the forward shock and entered its bow shock phase can be estimated as 
\begin{equation}\label{eq:released_t}
E_{\rm r} (t) = \int_{t_{\rm BS}}^{t} \!\! dt' \, \mathcal{L}_{e^\pm}(t') = \frac{I \Omega_0^2}{2}  \left[ \frac{1}{1 + t_{\rm BS} / \tau_0} - \frac{1}{1 + t / \tau_0} \right],
\end{equation}
assuming efficiency of order unity of conversion to pairs of the spin-down luminosity. For $t \rightarrow \infty$ we get an asymptotic energy of $\sim 8.6 \times 10^{47}$~erg for a pulsar having initial period $P_0 = 100$~ms, namely $\sim 40$\% of the total available energy, $\frac{1}{2}\Omega_0^2 I$, is released after the pulsar has escaped the parent SNR.

\begin{figure}[t]
\centering
\includegraphics[width=0.48\textwidth]{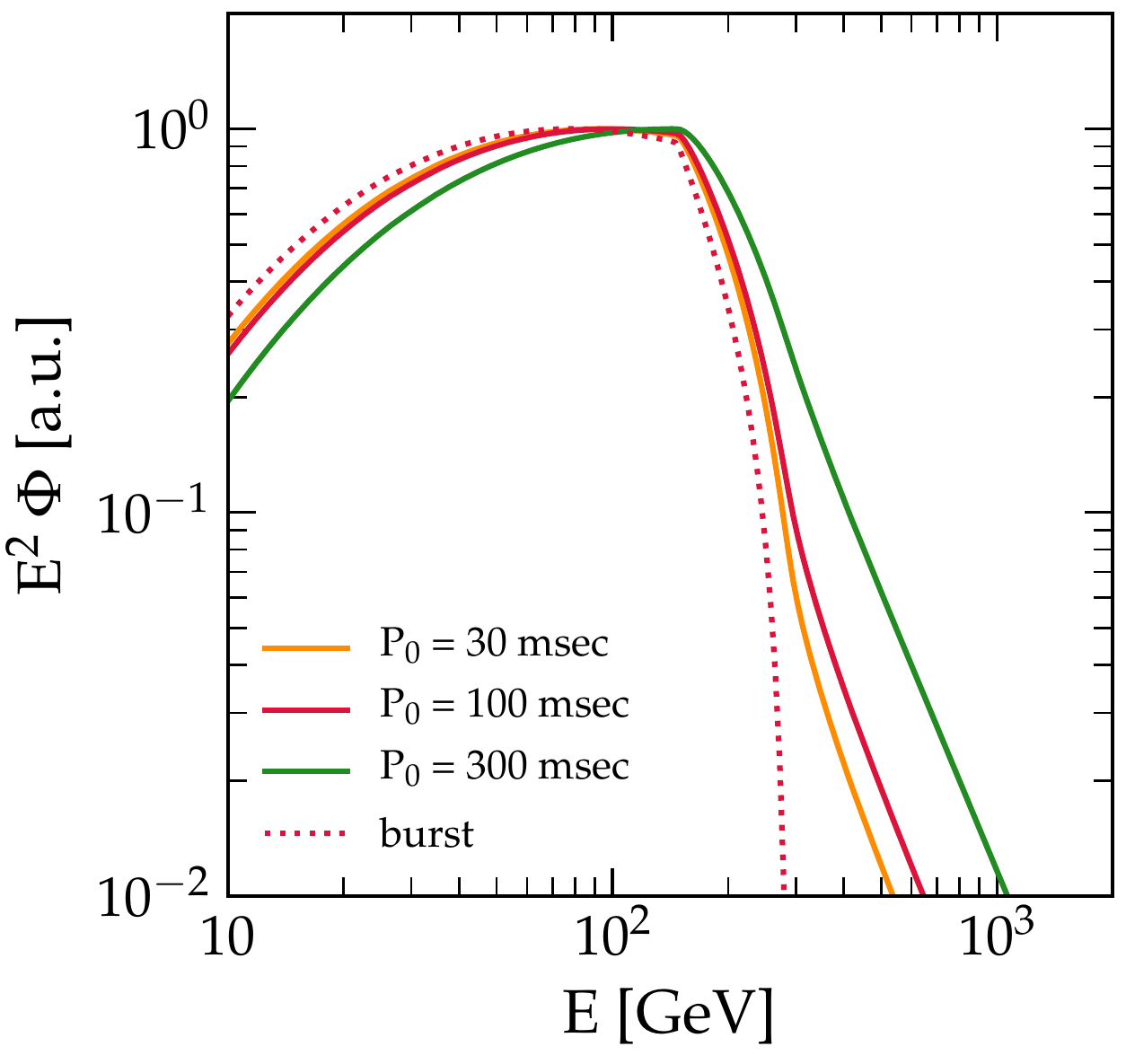}
\caption{The positron or electron flux (normalized to the maximum) computed from a single PWN at a distance of 3 kpc and with an age of 2 Myr for different values of the initial spin period $P_0$. The results obtained with the burst-injection approximation (notice that in the burst case the normalized spectrum is practically independent on $P_0$) are shown with a dotted line.} 
\label{fig:singlepulsar}
\end{figure}

Notice that for a pulsar with shorter initial period, say $P_0 = 30$~ms, the total energy available is higher since it is $\propto P_0^{-2}$ but the $\tau_0$ is shorter, so that the energy released is $1.3 \times 10^{48}$~erg which is only $5$\% of the total available energy associated to the pulsar spin down. 
In other words, faster pulsars have more energy available but they release most of it in particles that are confined inside the SNR. In principle the release of these particles at later times may contribute to the synthesis of the overall CR lepton spectrum in the Galaxy. In fact, however, this contribution turns out to be negligible, owing to adiabatic and radiative losses. The latter, in particular, become dramatic during the reverberation phase (see e.g.~\cite{Bandiera2020mnras} for an updated discussion of the PWN behaviour during this phase). A quantitative estimate of the effect of losses was given by~\cite{Gallant2018nppp} for a pulsar with an initial period of 40 ms, showing that, for typical values of the parameters, all the pairs released by the pulsar in the SNR would be degraded to energies lower than 50 GeV by the time of release in the ISM (see Fig.~2 of~\cite{Gallant2018nppp}). A completely analogous behaviour can be expected for a 30 ms pulsar, while compression can only be larger for a longer period pulsar, making losses even more severe. In addition, it should be noted that the estimate provided by~\cite{Gallant2018nppp} is likely a lower limit on the energy losses, because the compression considered in that work is actually only adiabatic: since the pressure in the nebula is carried entirely by relativistic pairs, the increased synchrotron losses during compression are likely to have a positive feedback, resulting in a radiative catastrophe. The overall energy content of the nebula is correspondingly degraded: in fact, most of the pulsar wind energy goes into particles with energy $E_b\approx 500$ GeV, which will be released in the ISM with at most a ten times lower energy. This means that even for the case of a 30 ms pulsar, that was depositing in the remnant 95\% of its energy before leaving, only few \% of the pulsar spin down energy $\sim \dot E \tau_0$ ends up in the form of pairs released in the ISM when the SNR finally dissipates. We neglect this contribution in the following.

In Fig.~\ref{fig:pulsar} we compare the temporal evolution of the luminosity $\mathcal{L}_{e^\pm}$, the energy released $E_{\rm r}$, and the cutoff energy $E_{\rm c}$, for three pulsars with $P_0 = 30$~ms,  $P_0 = 100$~ms, and $P_0 = 300$~ms, assuming $v_k = 280$~km/s and $\epsilon = 1$. 
The shaded area refers to the time before $t_{\rm BS}$, when the pulsar is inside the parent SNR. In addition to the points discussed above, one can appreciate how the maximum energy associated with the potential drop hardly falls below $\sim 100$ TeV, even at very long times. This has important implications in terms of modeling the diffuse gamma-ray emission recently detected from extended regions around some pulsars \cite{HAWC.2017.tevhalos}.

Finally, the contribution of the $i$-th source to the local flux of electrons or positrons is given by 
\begin{equation}
n^{\rm PWN}_i(t, E, \vec r) = \frac{1}{2} \frac{1}{b(E)} \int_{t_{\rm BS}}^{t-t_0} \!\! dt' b(E_*) \frac{Q_{\rm PWN}(E_s, t_*)}{\left[\pi \lambda_*^2\right]^{3/2}} 
\times \Sigma(x,y,\lambda_*) \chi(z,\lambda_*)
\end{equation}
where $E_*$ now satisfies $\Delta \tau(E, E_*) = t - t_0 - t'$ and $\lambda_* = \lambda_e(E, E_*)$.

In Fig.~\ref{fig:singlepulsar}, we illustrate the main qualitative differences between the burst solution of the transport equation and the case of continuous, time-dependent injection discussed above. The figure shows the flux from an individual source located at a distance from the Sun of $3$ kpc, an age of $2$ Myr, and with different initial periods $P_{0}-30,~100, and~300$ ms. The dotted line shows the burst solution while the other lines refer to the case of continuous injection. All curves are normalized so as to have the same peak. Hence, the burst case does not depend on $P_{0}$. Although there is an intrinsic cutoff in the injection of pairs due to the potential drop, this has little influence on the spectra at Earth which are dominated by energy losses. The main difference between the burst case and the case with time-dependent injection is in the high-energy behavior of the solution: the high-energy part of the spectrum at Earth is increasingly more populated when $P_{0}$ is larger, because the source injects particles into the ISM for longer times.  
We compare this result with the one obtained using the common assumption of a burstlike event, in which all positrons are released at $t_{\rm BS}$. In doing so, we compute the flux as in Eq.~\ref{eq:burstsolution}, with the source term in Eq.~\ref{eq:source-pwn} normalized to a total energy of 
\begin{equation}
E_{\rm r} (t_{\rm age}) = \int_{t_{\rm BS}}^\infty \!\! dt' \, \mathcal{L}_{e^\pm}(t') \ .
\end{equation}
The cutoff in energy is taken to be the potential drop at the bow shock formation. We notice however that this is not really relevant since the observed cutoff is mainly determined by the energy losses.

\begin{figure}
\centering
\includegraphics[width=0.48\columnwidth]{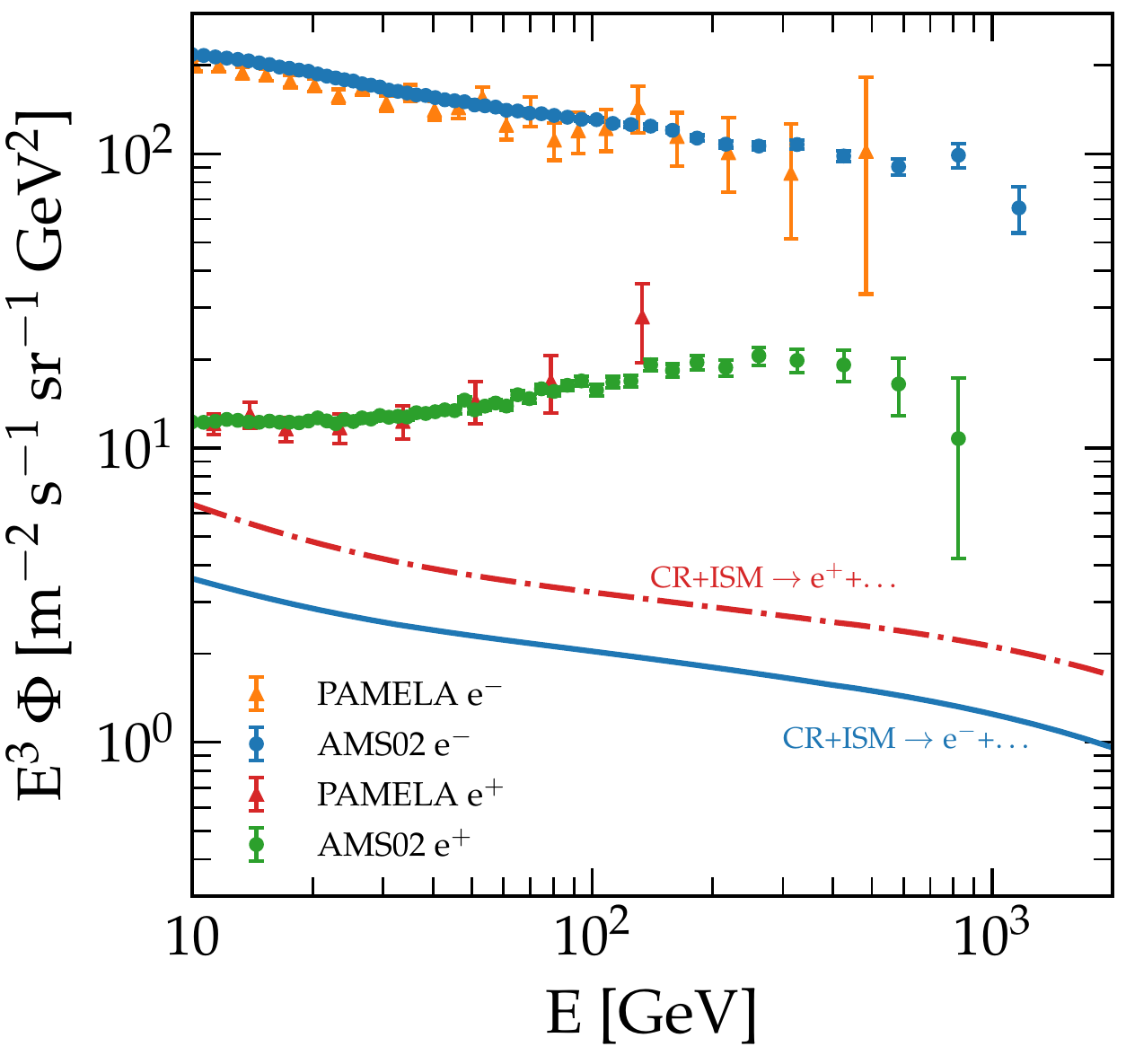}
\caption{The computed interstellar secondary positron (red dash-dotted curve) and electron (blue solid curve) spectra. Absolute measurements of the positron and electron fluxes by AMS-02~\cite{AMS02.2019.positrons,AMS02.2019.electrons} and PAMELA~\cite{PAMELA.2013.positrons,PAMELA.2011.electrons} are also shown.}
\label{fig:secondary}
\end{figure}

\subsection{Secondary leptons}
\label{sec:secondaryep}

Secondary electrons and positrons originate from the inelastic scattering of hadronic CRs (mainly protons and $\alpha$ particles) in the ISM. The main channel is the production of charged pions, although other processes also contribute. Given the positive charge of the parent CR particles, a slight excess of positrons over electrons is expected~\cite[see e.g.][]{Kamae2006apj}.

The steady-state source term for secondaries may be written in a general way as 
\begin{equation}
{\mathcal Q}_{e^\pm}(E_s, \vec r_s)  = 4 \pi \, f_{\rm ISM} n_{\rm H}(\vec r_s) \int_{E_s}^\infty dE_i \left[ \Phi_{\rm H}(E_i, \vec r_\odot) + 4^{2/3} \Phi_{\rm He}(E_i, \vec r_\odot) \right] \frac{d\sigma_{pp}^{\pm}}{dE}(E_i, E_s)
\end{equation}
where $d\sigma_{pp}^{\pm}/dE$ is the production differential cross section of $e^\pm$ by proton-proton collisions for which we make use of the parameterization provided in~\cite{Kamae2006apj}. The factor $4^{2/3}$ is derived by assuming that, for geometrical reasons, CR helium contributes to the production cross section as $A^{2/3}$ protons with $A$ the mass number.
These cross sections are still affected by some degree of uncertainty, in particular different approaches to model the contribution of helium to the total production cross sections can result in $\sim 30$\% differences as discussed in~\cite{Evoli2018jcap}.

In order to model the gas density as a function of position in the Galaxy, we adopt the distributions of neutral and molecular hydrogen provided by~\cite{Ferriere1998apj}, sum them to obtain $n_{\rm H}(\vec r_s)$, and apply the usual nuclear enhancement factor $f_{\rm ISM} \sim 1.4$ to take into account heavier elements in the target.

For the local interstellar spectra of protons, $\Phi_{\rm H}$, and helium, $\Phi_{\rm He}$, we adopt the parametrization given in~\cite{Boschini2017apj} based on various time-dependent measurements of the cosmic-ray flux at Earth and on a recent model of the solar modulation that they used to demodulate the observed flux. However, since we are mainly interested in secondaries of energy above $\sim 20$~GeV, produced by primaries with energy $\gtrsim 100$~GeV, solar modulation plays no significant role.  
In the calculations below we will assume that the nuclear CR flux is spatially uniform over a distance from the Sun comparable with the loss length $\lambda_e$. This assumption, that allows us to avoid using the Green function formalism for the primary nuclei, is theoretically justified by the absence of relevant losses for the nuclear component and observationally justified by the measured flat radial profile of the gamma-ray emission of hadronic origin~\cite{Grenier2015araa}. 

Under these assumptions, the contribution of secondary particles to the local CR flux can be computed as 
\begin{equation}\label{eq:secondaries}
\Phi^{\rm sec}(t_\odot, E, \vec r_\odot) =  \frac{f_{\rm ISM} c}{\pi^{3/2} b(E)} \int_E^\infty \!\!\! dE_s  I_V(E_s, E)
\int_{E_s}^\infty \!\! dE_i  \left[ \Phi_{\rm p}(E_i) + 4^{2/3} \Phi_{\rm He}(E_i) \right] \frac{d\sigma^\pm_{pp}}{dE}(E_s, E_i)
\end{equation}
where $I_V$ is proportional to the density averaged over a sphere of radius $\lambda_e$:
\begin{equation}
I_V[\lambda_e(E_s, E)] = \frac{1}{\lambda_e^3} \int_{\rm MW} \!\! dV \, n_{\rm H} (\vec r)   
\Sigma(x_\odot, y_\odot, \lambda_e) \chi(z_\odot, \lambda_e)
\end{equation}

Figure~\ref{fig:secondary} shows the computed secondary positron and electron spectra together with the fluxes measured by AMS-02 and PAMELA. One can appreciate how the secondary contribution to the positron flux is $\sim 50$\% at $\sim 10$~GeV, and rapidly decreases at higher energies, as expected. 
The secondary contribution to the observed electron flux is less than 2\% at $\sim 10$~GeV and even smaller at higher energies. 

\section{Source distribution and parameters}
\label{sec:spirals}

\begin{figure*}[t]
\hspace{\stretch{1}}
\includegraphics[height=0.45\textwidth]{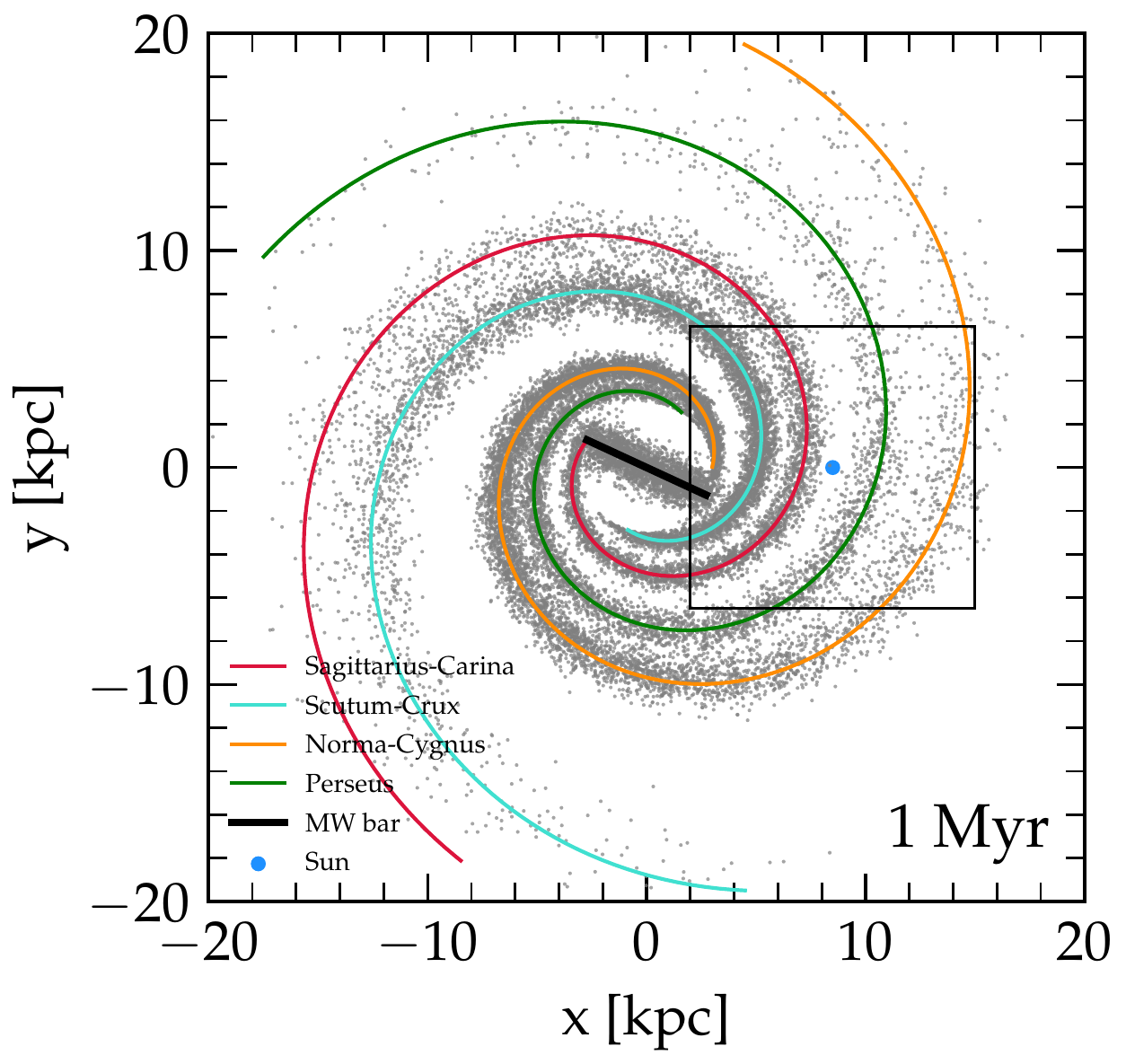}
\hspace{\stretch{1}}
\includegraphics[height=0.45\textwidth]{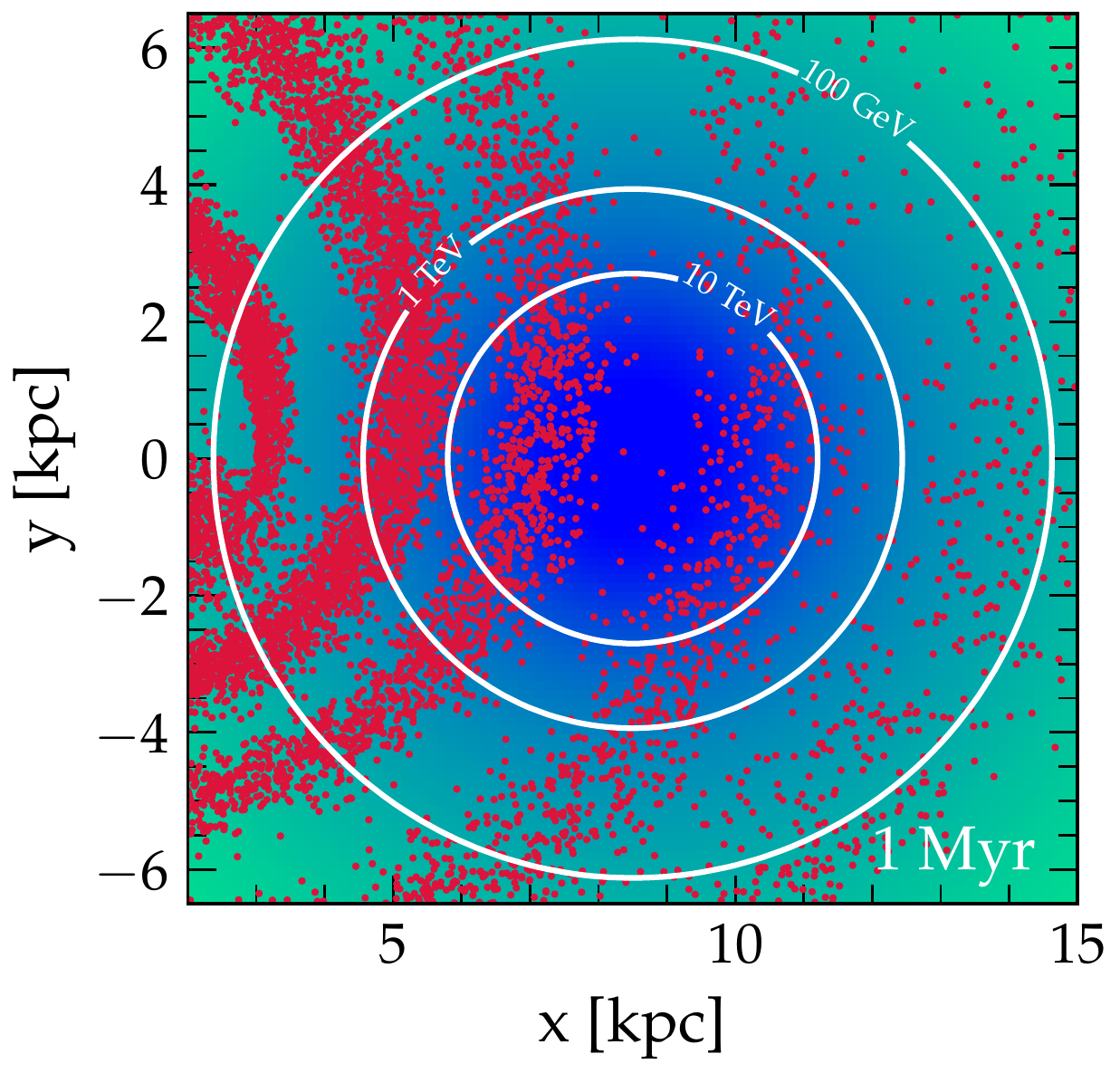}
\hspace{\stretch{1}}
\caption{The plot on the left shows the position of the explosions in the Galactic plane in a given realization and for a simulation time of 1 Myr. In the same plot we show the loci of the four arms of the Milky Way spiral structure. The position of the Sun is represented by the thick (blue) circle. 
The plot on the right shows an enlarged version of this plot (centered on the Sun location), where additionally we show the particle horizon for three different energies: 100 GeV, 1 TeV and 10 TeV.}
\label{fig:pattern}
\end{figure*}

\begin{figure}[t]
\centering
\includegraphics[height=0.48\textwidth]{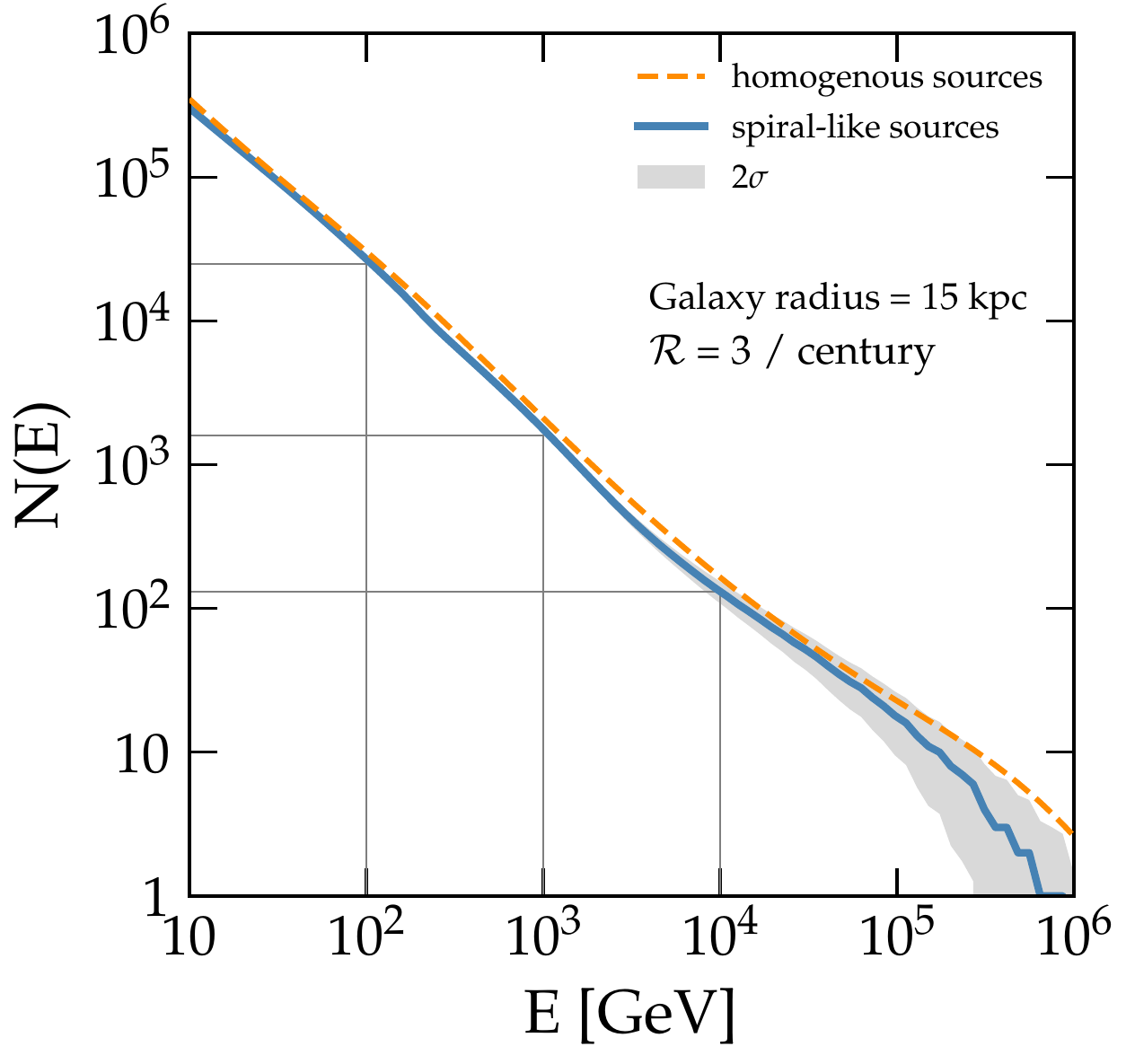}
\caption{The number of sources contributing to the local flux at a given energy $E$ for a Galactic source rate of $\mathcal R = 3$~event per century and spiral-arm distribution as described in Sec.~\ref{sec:spirals}. The solid blue line denotes the median and the gray band the 2$\sigma$ variation over $10^4$ realizations. The dashed orange line shows the expected result in case of homogeneous source distribution as in Eq.~\ref{eq:homo}.}
\label{fig:nsources}
\end{figure}

The spiral-arm structure of our Galaxy is of the utmost importance for the prediction of the flux of electrons and positrons observed at Earth. The peculiar position of the Solar System, which is located in an interarm region, is such that the distance to the closest arm is comparable with the distance traveled by a high-energy electron under the effect of radiative losses. Hence the spectrum and level of fluctuations for energies in the $\gtrsim$ TeV region are profoundly affected by the spiral structure. Similar considerations hold for the expected anisotropy. 

The stars that give rise to most supernova explosions are located in star-forming regions which in turn cluster inside the spiral arms of the Galaxy and in the Galactic bar. Explosions associated with older stars, such as type Ia SNe, are more spread in the ISM but still concentrated mainly inside the arms. Their rate is about 1/3 of that of core collapse SNe although the energetics of the two are rather similar, $\sim 10^{51}$ erg. 

A variety of models describing the spiral-arm structure of the Galaxy have been put forward, and most of them consist of four major spiral arms (see, e.g.~\cite{FaucherGiguere2006apj,Vallee2017astrv,Steppa2020aa}). We adopt the logarithmic spiral-arm parametrization proposed by~\cite{SteimanCameron2010apj}, which was derived from observations of FIR cooling lines, [CII] and [NII], of the ISM. These lines trace increased gas density and UV radiation fields and are therefore thought to mark the presence of star formation regions. 
The existence of a central bar is inferred from the evidence of a strong asymmetry in the number of stars with respect to the direction of the Galactic Center, although the detailed structure of the bar is still a matter of debate. In fact, several authors advocate a picture in which the Milky Way might feature two or more distinct bars~\cite{Wegg2015mnras}.
For the sake of simplicity, we model the central bar as in~\cite{Churchwell2009} with a half-length of $\sim 3.1$~kpc and an angle $\theta = 20^{\rm o}$ with respect to the Galactic Center-Sun line. The aspect ratio is found to be roughly 10:4:3 (length:width:height), making this structure much more vertically extended than the thin stellar disk.
The procedure we adopt to assign a location to each SN event is the following: we first choose at random the galactocentric distance of the source, $\tilde r$, from a distribution that is proportional to the functional form suggested by~\cite{Lorimer2006mnras} based on the Parkes multibeam survey at 1.4~GHz~(model \emph{C} in their Table 7).

If $\tilde r$ is smaller than 3 kpc, we assume that the event is associated with the Galactic bar and we assign an initial position ($x_0, y_0$) along it. We then spread the ($x$, $y$) coordinates of that SN by translating it by a distance drawn from a normal distribution centred at zero with standard deviation $\sim 1$~kpc, chosen so as to roughly resemble the observed aspect ratio. Furthermore, in this case SNe have a distance $z$ away from the disk that follows a Gaussian distribution with a scale height of $\sim 1$~kpc  and mean of 0 pc.

If $\tilde r$ is larger than 3 kpc we choose the SN locations so that their projections lie on arms and subsequently we alter them to simulate a spread about the arm centroids. Specifically, we first draw a random integer number between 1 and 4 assuming that the birth rate is proportional to the [CII] emissivity of each arm (see Table 4 in~\cite{SteimanCameron2010apj}). This number identifies the arm in which the SN is located (Carina-Sagittarius, Crux-Scutum, Perseus or Norma-Cygnus, as in Table 3 in~\cite{SteimanCameron2010apj}). 
At this point we choose at random the position of the source by drawing the coordinates $\tilde \phi$, and $\tilde z$ from a distribution that is proportional to the emissivity function in Eq.~9 of~\cite{SteimanCameron2010apj}. 
If not differently stated, SNe are generated at the canonical rate $\mathcal R = 3$ per century. We discuss the impact of changing this value in the next section.

In Fig.~\ref{fig:pattern} we show the location of the sources in a given realization of their distribution, for a simulation time of 1 Myr. 
The Sun is located at $R_\odot = 8.5$~kpc along the positive $x$ axis.  
In the right panel of the same figure we enlarge the source spatial distribution for a closer look at the vicinity of the Sun and we overplot the circles with radius $\lambda_e$ corresponding to energies of 100 GeV, 1 TeV and 10 TeV. 
This plot provides already a first visual assessment of the decreasing number of sources able to contribute to the local flux as the electron energy increases.

More quantitatively, the sources that can contribute to the flux at Earth at a given energy $E$ are those lying within a distance from Earth such that the propagation time is shorter than the loss time at that energy.
In case of uniformly distributed sources in the disk of the Galaxy, with radius $R_g$, the number of sources within a distance $\lambda_e$ from Earth, exploding in a loss timescale $\tau_{\rm loss}$ would be 
\begin{equation}\label{eq:homo}
N (E) \sim \frac{\mathcal R \tau_{\rm loss}(E) \lambda_e^2(E)}{R_g^2} .
\end{equation}

Since both $\tau_{\rm loss}$ and $\lambda_e^2$ are decreasing functions of energy, this simple estimate shows that the number of contributing sources is a rapidly decreasing function of energy. For $R_g = 15$~kpc, Eq. \ref{eq:homo} gives $\sim 2 \times 10^4$ sources contributing at 100 GeV and $\sim 1.5 \times 10^3$ at 1 TeV. At 10 TeV only about $\sim 100$ sources can contribute. These numbers are considerably larger than those reported in previous studies. The main reason for this difference is that we adopted a diffusion coefficient that was derived in recent studies of the AMS-02 data on primary and secondary nuclei \cite{Evoli2019prd}, which implied an energy dependence at low energies ($\delta = 0.54$) faster than in previous studies (for instance $\delta = 0.33$ was assumed in \cite{BlasiAmato2011,Mertsch2011jcap}). Moreover, the analysis of the beryllium flux observed by AMS-02 led the authors of Ref. \cite{Evoli2020prd} to conclude that the size of the Galactic halo should exceed $\sim 5$ kpc, larger than previously used. As a consequence the normalization of the diffusion coefficient is also required to be larger, thereby making the effective volume where sources can contribute to the lepton flux larger.

To compare these numbers with a realistic case, we draw $10^4$ realizations of the spatial distribution of $\mathcal R \times 100$ Myr sources and for each of them we count as a function of energy $E$ the number of objects that lie within the horizon $\lambda_e(E)$ and with an age $t_a$ smaller than $\tau_{\rm loss}(E)$. The median of the cumulative distribution functions is shown in Fig.~\ref{fig:nsources} and compared with the result given by Eq.~\ref{eq:homo}.
We notice that the two curves deviate from one another starting at E $\sim$ 10 TeV where the presence of a spiral structure becomes important and the fluctuations associated with different realizations of the source distribution become sizeable. However, as long as energies below 1 TeV are considered, the inclusion of spiral arms does not impact the number of contributing source.

Isolated neutron stars are thought to be formed in core-collapse SN events, which are a fraction of the total SNe in our Galaxy. In our analysis, we assume that pulsars are formed in 80\% of total SN explosions, chosen randomly for each realization.
The initial spin period $P_0$ of each pulsar is chosen from a normal distribution with mean $\langle P_0 \rangle$ and standard deviation $\sigma_{P_0}$. In order to ensure that only positive values of $P_0$ are drawn we assume that the distribution function is 0 for $P_0 < 0$.

\begin{figure}[t]
\centering
\includegraphics[height=0.48\textwidth]{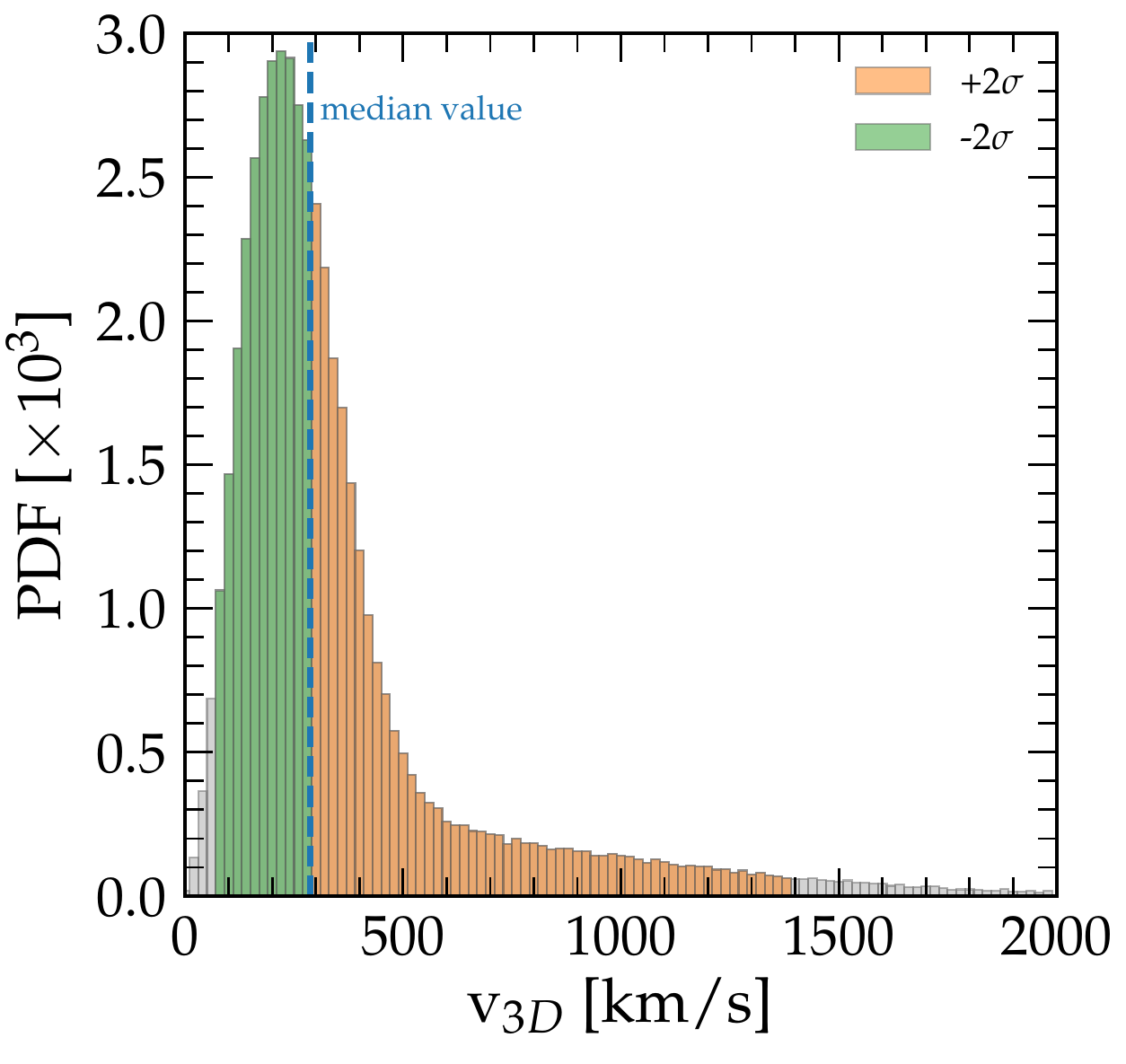}
\caption{Distribution of pulsar kick velocities calculated following~\cite{FaucherGiguere2006apj}. The vertical dashed line identifies the median kick velocity.}
\label{fig:vkdistro}
\end{figure}

Following~\cite{FaucherGiguere2006apj}, the pulsar birth velocity distribution is taken as the sum of two Gaussians (see their Eq.~7) for each of the 3 velocity components. 
Correspondingly, the three-dimensional velocity distribution is Maxwellian, as shown in Fig.~\ref{fig:vkdistro}. For the parameters provided in~\cite{FaucherGiguere2006apj}, the median value of the velocity norm is $\sim 300$ km/s and 99\% of the pulsars have kick velocity between 75 and 1400~km/s.

\section{Results}
\label{sec:results}

\subsection{Primary electrons from SNRs}
\label{sec:electrons}

\begin{figure*}[t]
\hspace{\stretch{1}}
\includegraphics[width=0.47\textwidth]{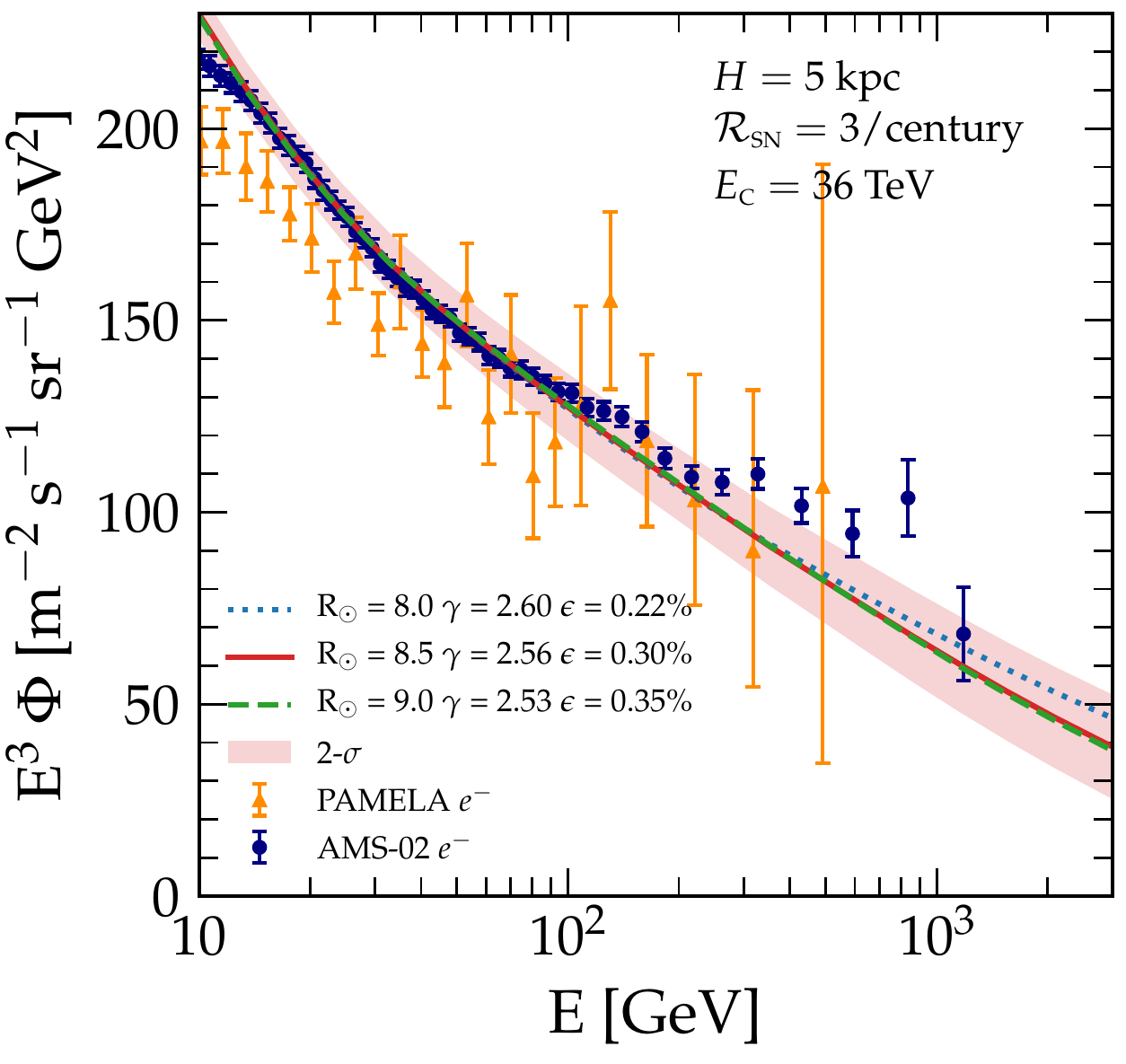}
\hspace{\stretch{1}}
\includegraphics[width=0.47\textwidth]{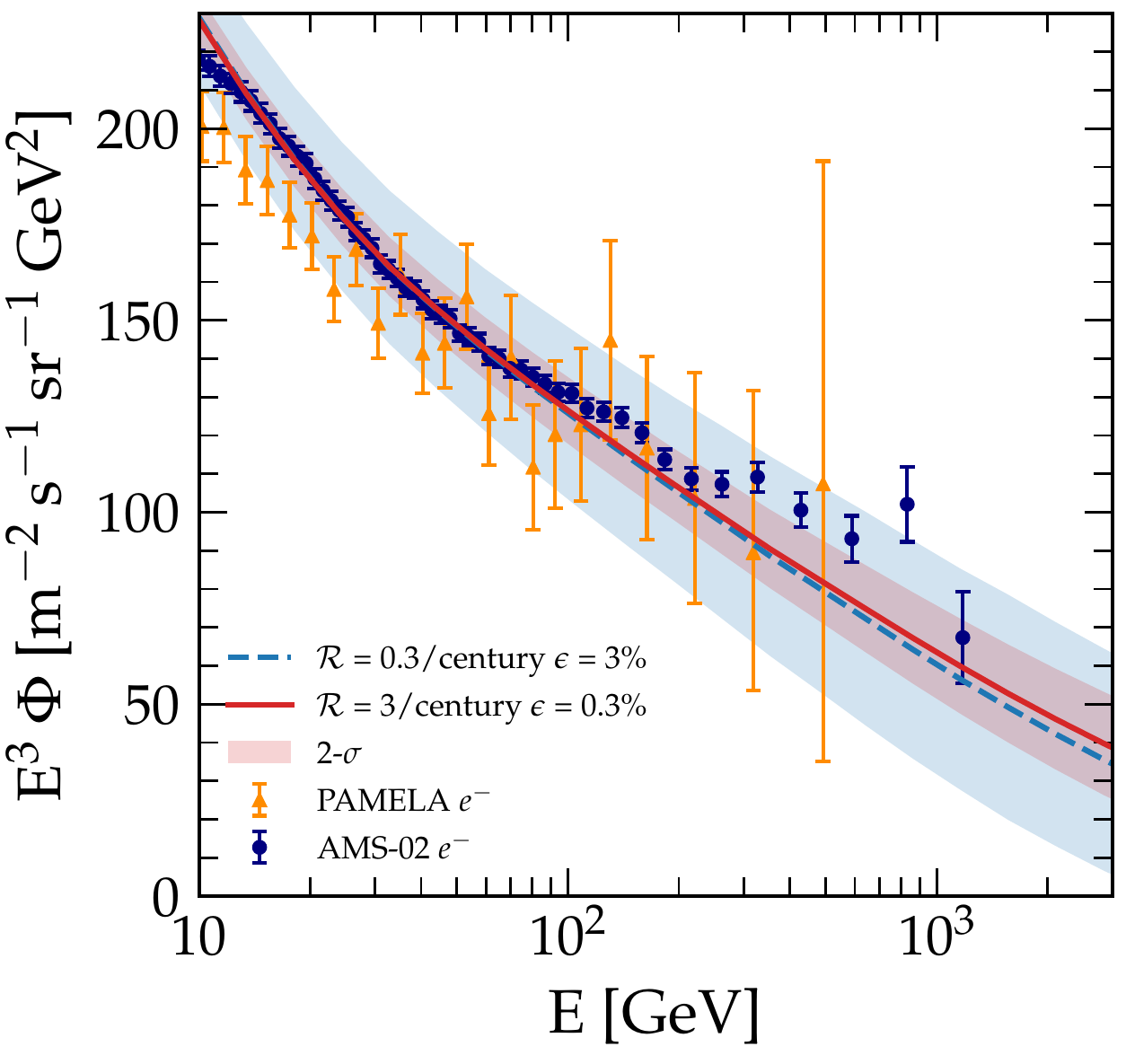}
\hspace{\stretch{1}}
\caption{The spectrum of cosmic-ray electrons resulting from the sum of all SNRs throughout the Galaxy. Secondary electrons as calculated in Eq.~\ref{eq:secondaries} are also included. Lines represent the median flux computed over $10^4$ realizations of the Milky Way and the shadowed areas represent the 2$\sigma$ fluctuations above and below the median value. In the left panel, we show the flux computed for three different positions of the observer, $R_\odot = 8$ (dotted blue curve), $8.5$ (solid red curve) and $9$~kpc (dashed green curve). We modified accordingly the injection spectrum, both in slope and efficiency, to best reproduce the data above 20 GeV. 
In the right panel, we compare the case with fiducial injection rate $\mathcal R = 3$~SN per century (red solid curve) with the case where the SNR rate is assumed to be a factor of 10 smaller (blue dashed curve). In the latter case, the efficiency required is about 10 times larger. The measurements of PAMELA~\cite{PAMELA.2011.electrons} (orange triangles) and of AMS-02~\cite{AMS02.2019.electrons} (blue dots) are also shown.}
\label{fig:electronsSNR}
\end{figure*}

In this section we compare the results of our calculations with the measurements of the electron flux obtained by PAMELA~\cite{PAMELA.2011.electrons} and AMS-02~\cite{AMS02.2019.electrons}. In order to discuss individual effects, here we focus on the contribution to the electron flux due to SNRs and secondary production, while a fit including the contribution of pulsars will be discussed later in this article. As discussed in Sec.~\ref{sec:secondaryep}, the contribution of secondary production to the electron flux is small, but for completeness it is included here. 

The calculations are performed by generating $N=10^4$ Monte Carlo realizations of our Galaxy as described in Sec.~\ref{sec:spirals}. Locations and times of occurrence of SNRs are generated at random over a time span that is long enough to ensure that a stationary solution is reached at energies $\gtrsim 10$ GeV. A duration of $T = 100$~Myr is more than sufficient to guarantee that this is the case, since it well exceeds the timescale for energy losses $\tau_{\rm loss}$ at $10-20$ GeV. 

For each realization the total flux at Earth is computed by adding the contributions of individual sources as written in Eq.~\ref{eq:burstsolution}. As in~\cite{Mertsch2011jcap}, we exclude sources that are not causally connected with Earth location, namely sources that, given their age and location, cannot contribute to the flux at the Earth without superluminal particle motion. Given the realizations, we evaluate the median flux at each energy, while the fluctuations are computed with respect to the median (also known as median absolute deviation).

The flux of electrons (SNR plus secondary electrons) at the Earth location is shown as a red solid curve in the left panel of Fig.~\ref{fig:electronsSNR} for the nominal position of the Sun ($R_\odot = 8.5$~kpc) and for the nominal rate of SNRs in the Galaxy ($\mathcal R = 3$ per century).
Fluctuations at 2$\sigma$ are shown with a red shadow band.

In order to fit the observed electron spectrum, primary electrons must be injected with a steeper spectrum than protons: $\gamma_{\rm e} \sim 2.56$, while $\gamma_{\rm H} \sim 2.3$~\cite{Evoli2019prd} (and we will show in the next section that the injection electron spectrum is even steeper after accounting for the pulsar contribution). It has been recently proposed that a different source spectrum for electrons and protons may be due to radiative losses of leptons in the downstream region of a SNR shock \cite{Diesing2019prl}. This proposal relies on the possibility to have relatively large magnetic fields in the late stages of the shock evolution, a scenario that requires further investigation.

Due to energy losses, the electron spectrum is rather sensitive to the average distance from sources: for instance the flux may change somewhat by slightly changing the average distance to the closest arm (different models provide estimates that differ by 1-2 kpc). Moreover there is some ambiguity about the presence of an armlet in between major arms, which might again provide a few local sources relevant for the high-energy electron spectrum. In order to provide an estimate of the importance of these effects we repeat the calculations illustrated above but moving the position of the Sun by 500 pc inward and outward with respect to the reference value $R_\odot = 8.5$~kpc. The results are shown in the left panel of Fig.~\ref{fig:electronsSNR} (blue dotted and green dashed lines as labeled). 
It is worth stressing that: (i) the main differences appear at $E\gtrsim 500$ GeV, as expected; (ii) when the Sun is positioned farther away from the closest spiral arm, the mean distance to the closest source is, on average, larger. Hence the injection spectrum in this configuration required to fit the data is a bit harder than for the canonical position. The opposite happens if the Sun is positioned closer to the nearest spiral arm. (iii) Finally, the efficiency of conversion of SN energy to CR electrons increases slightly while positioning the Sun farther away from the nearest spiral arm. However such an efficiency remains of order $\sim 0.3\%$ for the canonical rate of SN explosions in the Galaxy. 

When the rate of SN explosions is artificially reduced, the efficiency increases proportionally, and the fluctuations around the median increase, while the median is not substantially effected, with the exception of the energies $\gtrsim 1$ TeV. The injection spectrum necessary to fit the data is also independent of the rate of SN explosions, as expected. This is clearly illustrated in the right panel of Fig.~\ref{fig:electronsSNR}. 

\subsection{Positrons from PWNe}
\label{sec:positrons}

As discussed in Sec.~\ref{sec:pulsarmodel} and in Sec.~\ref{sec:spirals}, in our calculations pulsars are assumed to be produced in $80\%$ of SN explosions and to behave as continuous injectors of an equal number of electrons and positrons with a given injection spectrum, shaped as in Eq.~\ref{eq:injpulsar}. The birth kick velocity is generated at random from the distribution in Fig.~\ref{fig:vkdistro} and the corresponding escape time from the parent SNR is calculated accordingly. This allows us to compute the residual spin-down energy that can still be converted into pairs when the pulsar is in its bow shock nebular phase. 
The efficiency is calculated as the fraction of this energy that is to be converted to pairs in order to fit the positron spectrum observed by AMS-02~\cite{AMS02.2019.positrons}. 

\begin{figure*}[t]
\hspace{\stretch{1}}
\includegraphics[width=0.47\textwidth]{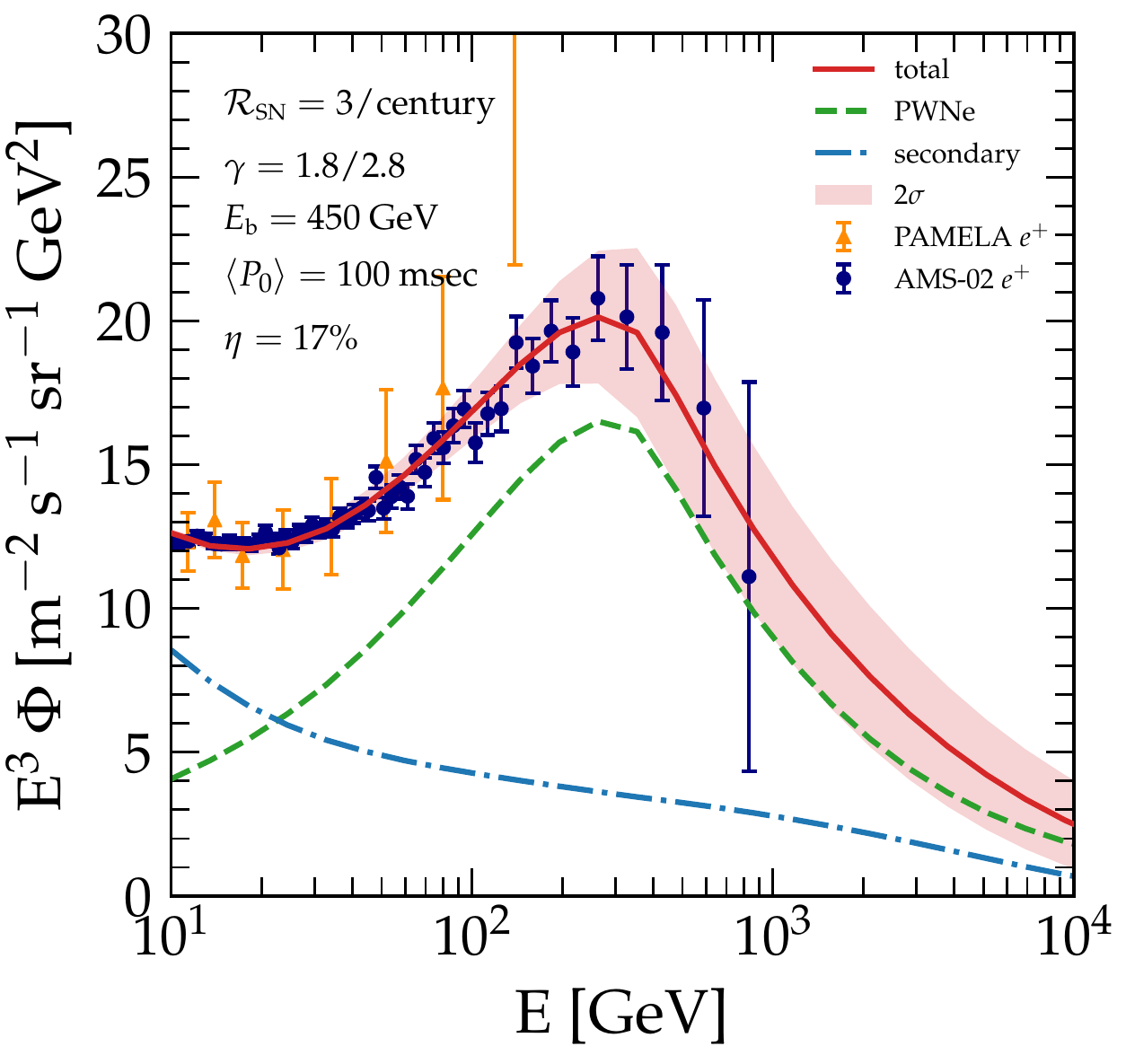}
\hspace{\stretch{1}}
\includegraphics[width=0.463\textwidth]{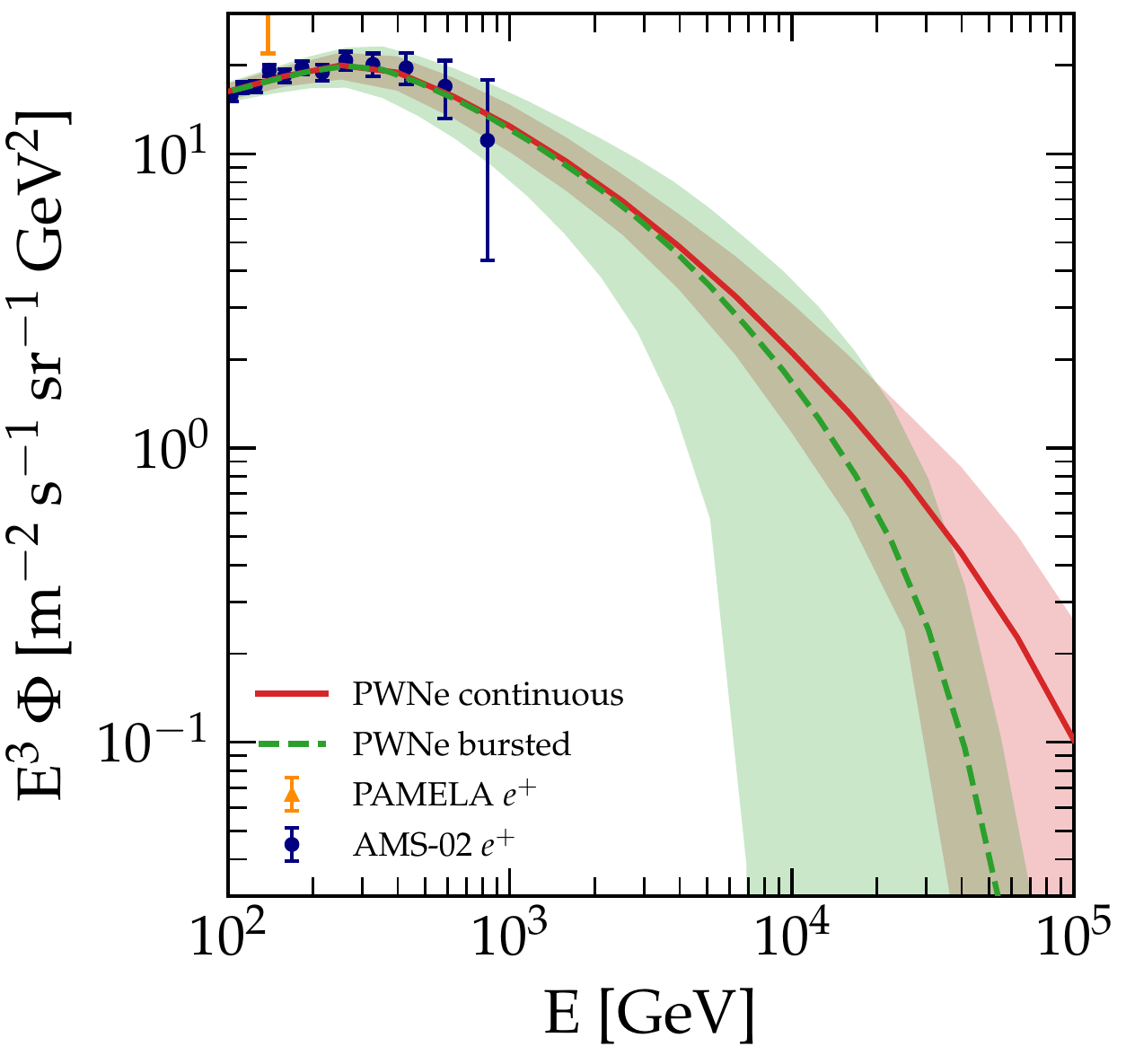}
\hspace{\stretch{1}}
\caption{Left panel: The spectrum of positrons resulting from the sum of all PWNe throughout the Galaxy assuming $\langle P_0 \rangle = 100$~ms is shown as a dashed green line. The injection spectrum is assumed to be a broken power law with slopes $1.8(2.8)$ below(above) the break at $E_b = 450$~GeV. Secondary positrons as calculated in Eq.~\ref{eq:secondaries} are also included and showed separately by a dash-dotted blue line. The total spectrum is shown with a solid red line. Right panel: The solid red line shows the total positron spectrum (the same as in the left panel); the latter is compared with the burst-injection approximation with the same assumed efficiency, $\eta = 17$\%, showed as a dashed green line. In both panels, the shaded areas represent 2$\sigma$ fluctuations around the median of the different realizations.}
\label{fig:positronsPWN}
\end{figure*}

The flux of positrons resulting from the superposition of pulsars generated stochastically following the spiral structure of the Galaxy is shown in the left panel of Fig.~\ref{fig:positronsPWN}, where we assumed $\langle P_0 \rangle = 100$~ms with a standard deviation of $\sigma_{P_0} = 50$~ms~\cite{Watters2011apj}. The red solid line shows the median flux of positrons as due to pulsars and secondary positrons from CR interactions (blue dash-dotted line). The pair spectrum injected by BSNe is required to have slopes $1.8\ (2.8)$ below (above) the break at $E_b = 450$~GeV. While the position of the break very well agrees with what deduced from observations of nonthermal emission from PWNe~\cite{Bucciantini2011mnras}, the spectrum appears to be, at all energies, somewhat steeper than typically inferred for well studied PWNe and BSNe~\cite{Kargaltsev2008aipc}. This difference might suggest that the particle population released by BSNe suffers non-negligible energy losses before leaving the source surroundings. 
The shaded area in Fig.~\ref{fig:positronsPWN} illustrates the role of fluctuations. Notice that at $\sim 10$ GeV about half of the positrons result from CR interactions in the ISM, while this contribution becomes smaller at higher energies. Because of this, the spectral feature observed in the electron spectrum and attributed to the onset of KN effects on the UV photons is hard to spot, although clearly this phenomenon occurs for positrons as well. The spectrum of positrons is described in an excellent way by the mixed contribution of positrons from pulsars and from CR interactions, the latter being strongly constrained by other observables, such as secondary nuclei (boron and beryllium). Notice that the break in the propagated positron spectrum is visible at an energy $E \sim 300$~GeV, lower than the break at the injection, because of energy losses.

The mean value of the initial period $P_{0}$ depends on modeling the observed population of existing pulsars and different approaches lead to somewhat different values of this parameter.
From the point of view of the production of pairs, the main difference when adopting different values of the initial period is in the efficiency of conversion of spin-down luminosity to pairs, which is much smaller for smaller values of $\langle P_0 \rangle$. We additionally test the case with $\langle P_0 \rangle = 300$~ms, which can be considered as the highest value found in the literature~\cite{FaucherGiguere2006apj}, and we obtain that the efficiency required in this case is $\eta = 85$\%.
 
One of the novelties of the calculations presented here with respect to previous literature is the fact that the time-dependent injection of pairs by pulsars is properly taken into account, rather than considering impulsive release by these sources (see Sec.~\ref{sec:pulsarmodel}). It is easy to understand that this different approach makes a difference especially at very high energies, where the duration of the source injection (although this concept requires a more careful definition when the injection rate is time dependent) becomes comparable with the propagation time from the source to the Earth. The burst approximation is increasingly better justified for distant sources and for smaller values of $P_{0}$. In the right panel of Fig.~\ref{fig:positronsPWN} we show the comparison between the time dependent prediction (red solid line) and the one based on the burst assumption (green dashed line), each with the associated estimate of the role of fluctuations (shaded areas). The predictions for the two approaches start departing from each other only at energies $\gtrsim 10$ TeV, while no appreciable difference can be noticed in the energy range where AMS-02 measurements are available. Both approaches lead to an excellent description of the measured positron flux at Earth. 

\begin{figure*}[t]
\hspace{\stretch{1}}
\includegraphics[width=0.47\textwidth]{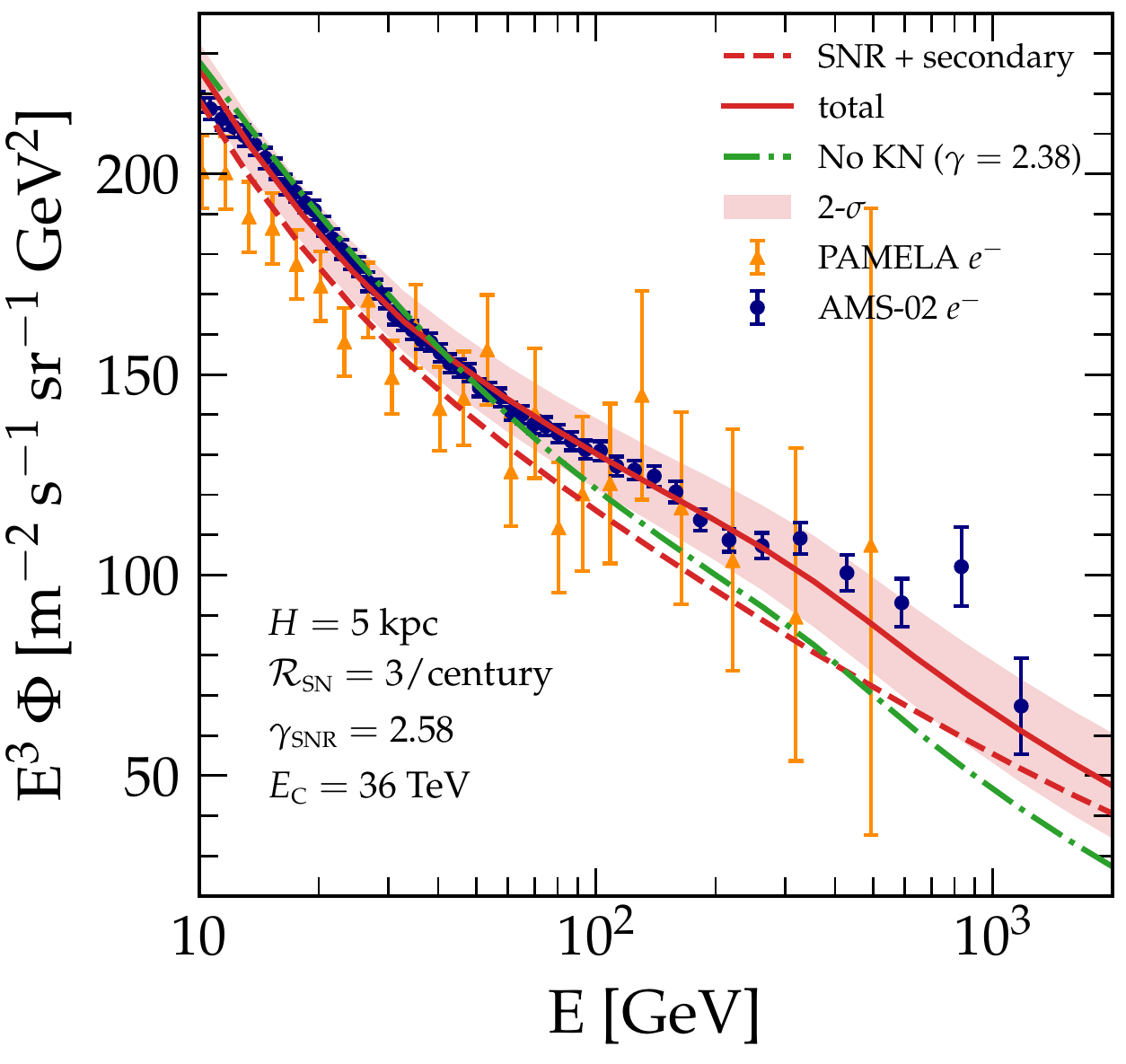}
\hspace{\stretch{1}}
\includegraphics[width=0.47\textwidth]{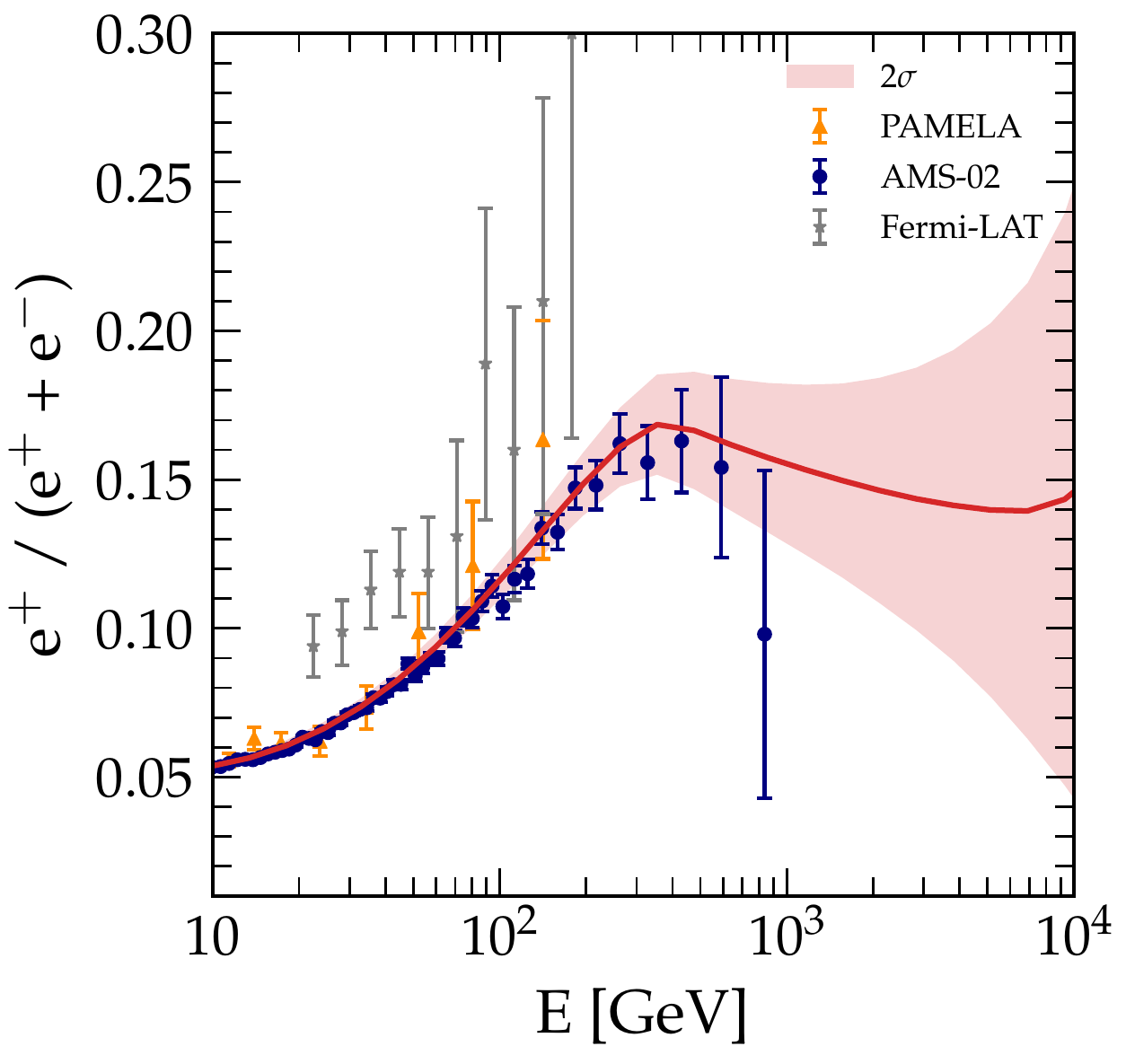}
\hspace{\stretch{1}}
\caption{Left panel: The spectrum of electrons from SNRs is shown with a red dashed line, and the total flux of electrons (including pulsars and secondary products of CR interactions) with a red solid line. The total flux obtained neglecting the KN effect (dash-dotted green line) is also shown. Right panel: positron fraction compared with data from PAMELA~\cite{PAMELA.2009.posfraction}, Fermi-LAT~\cite{FERMI.2012.posfraction} and AMS-02~\cite{AMS02.2019.electrons}. In both panels the shaded area shows the effect of fluctuations due to the stochastic nature of sources.}
\label{fig:electrons}
\end{figure*}

\subsection{Electron spectrum and positron fraction}

Once the contribution of pulsars to the positron flux is properly calibrated to the AMS-02 observations, as discussed in Sec.~\ref{sec:positrons}, we can proceed to the final part of our analysis, namely the calculation of the positron fraction. The first step in this direction is the reassessment of the spectrum of electrons. Compared with the results presented in Sec.~\ref{sec:electrons} we can now include the contribution of pulsars to the electron spectrum. This leads to a slight change in the parameters needed to fit the AMS-02 data, as we show in the left panel of Fig.~\ref{fig:electrons}: the dashed line shows the contribution of SNRs, which is now very close to the data, but the contribution of pulsars is crucial to fit the total spectrum of electrons. The flux of electrons as due to CR interactions with the ISM is small, but included in Fig.~\ref{fig:electrons}. The inclusion of the contribution of pulsars to the electron spectrum results in the need for a somewhat steeper injection spectrum in SNRs (slope 2.58 versus 2.56). 
Since the electron flux is dominated by the contribution of SNRs, the feature at $\gtrsim 50$ GeV due to the onset of KN on the UV background is clearly visible even after adding pulsars~\cite{Evoli2020prl}.
The dash-dotted green curve in Fig.~\ref{fig:electrons} illustrates the result of our calculations if we artificially remove the KN effect on the energy loss rate so that $b(E) \propto E^2$ for each of the ICS channels and for synchrotron losses. This case requires a harder injection spectrum to reproduce the low-energy data but is clearly unable to reproduce the electron measurements over the entire energy range, even after accounting for the contribution of PWNe.

The positron fraction is shown in the right panel of Fig.~\ref{fig:electrons}. The fluctuations at two standard deviations as due to the stochastic nature of the sources (shaded area) are also shown and compared with the ratio as measured by PAMELA \cite{PAMELA.2009.posfraction}, Fermi-LAT \cite{FERMI.2012.posfraction} and AMS-02 \cite{AMS02.2019.electrons}. The rising positron fraction is naturally reproduced by the pulsar contribution to the positron flux. At energies of a few hundred GeV, where the spectrum of positrons from pulsars becomes steeper, the fraction starts declining slightly. However, since the cutoff associated with the potential drop of pulsars is typically at much higher energy than the maximum energy of electrons accelerated in SNRs, for $E\gtrsim 5$ TeV the predicted positron fraction shows an uprise. On the other hand at the same energies fluctuations due to the stochastic nature of sources become large and the positron fraction at these energies is expected to show a rather irregular trend. Current data at high energies, as shown in the right panel of Fig.~\ref{fig:electrons}, have insufficient accuracy to clearly highlight these effects. 

\begin{figure}[t]
\hspace{\stretch{1}}
\includegraphics[width=0.47\textwidth]{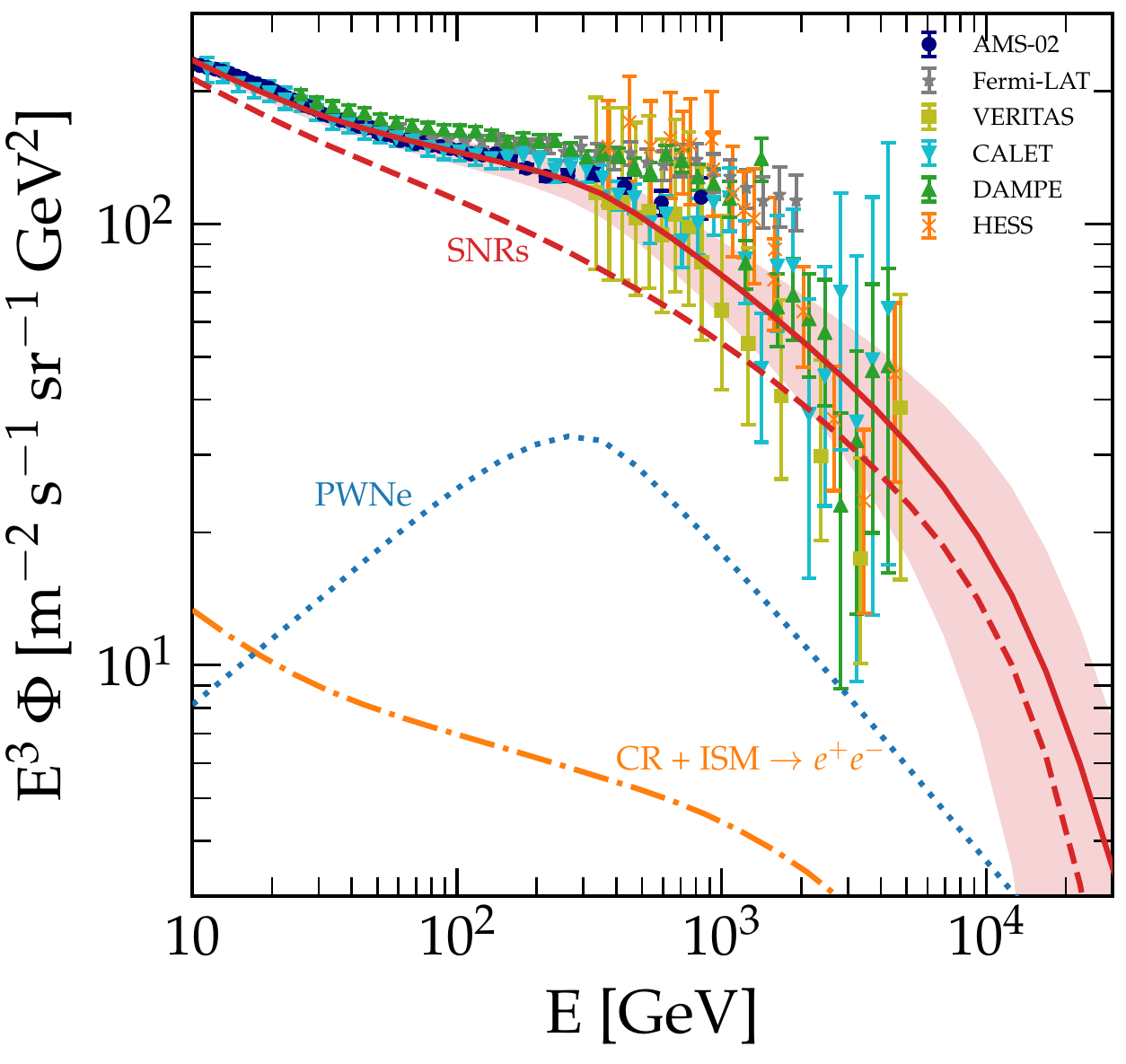}
\hspace{\stretch{1}}
\includegraphics[width=0.47\textwidth]{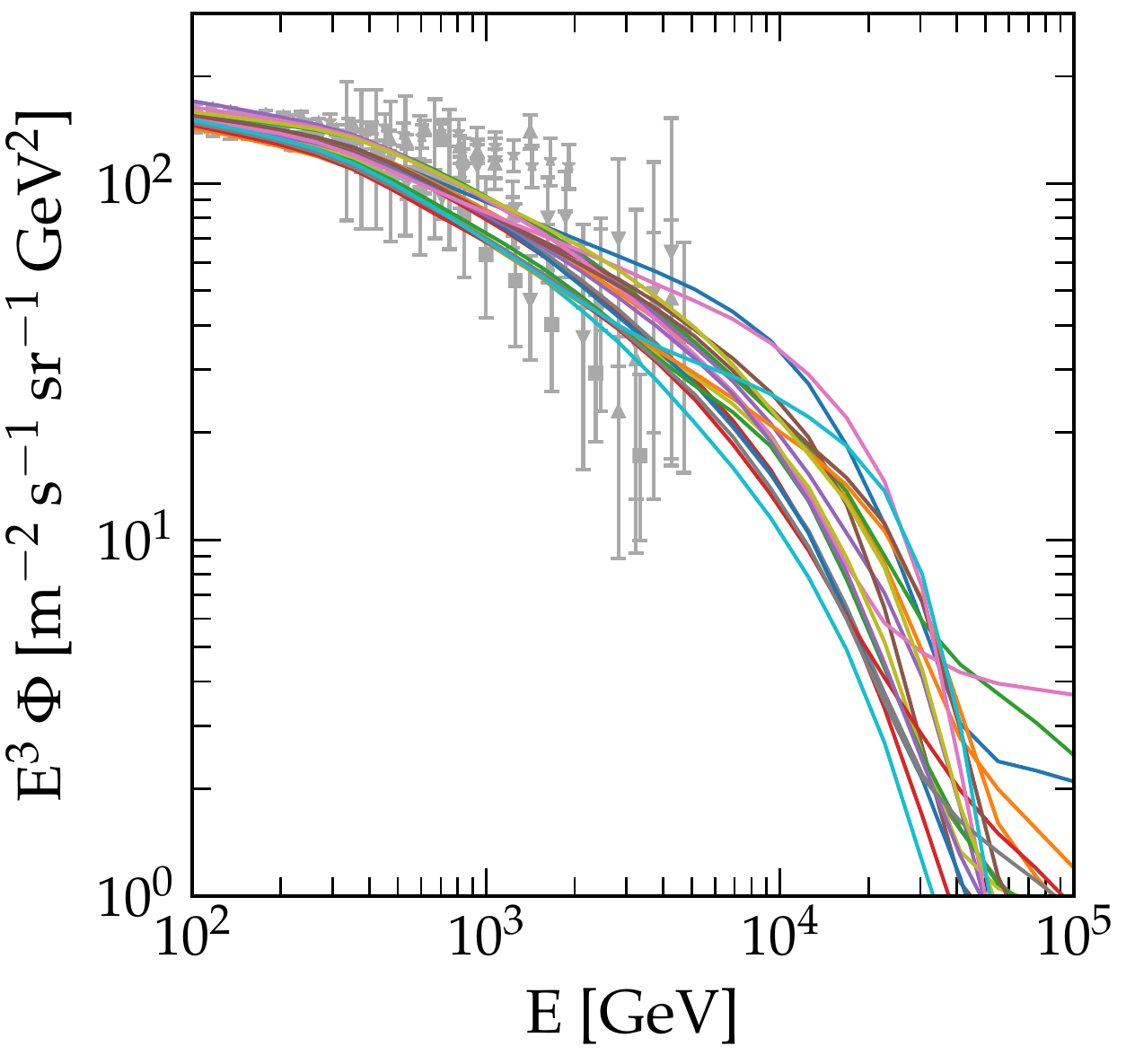}
\hspace{\stretch{1}}
\caption{Left panel: spectrum of $e^{+}+e^{-}$ from SNRs (dashed red line, only electrons), from pulsars (dotted blue line), from CR interactions in the ISM (dash-dotted orange line). The total flux (solid red line) is compared with available data from AMS-02~\cite{AMS02.2019.electrons}, CALET~\cite{CALET.2017.leptons}, DAMPE~\cite{DAMPE.2017.leptons}, HESS~\cite{HESS.2008.leptons,HESS.2009.leptons}, FERMI~\cite{FERMI.2017.leptons}, PAMELA~\cite{PAMELA.2011.electrons}, and VERITAS~\cite{VERITAS.2018.leptons}. The shaded area shows the effect of fluctuations (two standard deviations). Right panel: The colored lines show 20 different random realizations from the Monte Carlo simulations.}
\label{fig:leptons}
\end{figure}

The spectrum of $e^{+}+e^{-}$ as derived here and used to calculate the positron fraction is shown in Fig.~\ref{fig:leptons} (red solid curve) together with the corresponding data from AMS-02~\cite{AMS02.2019.electrons}, CALET~\cite{CALET.2017.leptons}, DAMPE~\cite{DAMPE.2017.leptons}, HESS~\footnote{The systematic errors for H.E.S.S. measurements are computed from two tables provided by the collaboration at~\url{https://www.mpi-hd.mpg.de/hfm/HESS/pages/publications/auxiliary/auxinfo_electrons.html} and \url{https://www.mpi-hd.mpg.de/hfm/HESS/pages/publications/auxiliary/auxinfo_electrons2.html}.}~\cite{HESS.2008.leptons,HESS.2009.leptons}, FERMI~\cite{FERMI.2017.leptons}, PAMELA~\cite{PAMELA.2011.electrons}, and VERITAS~\cite{VERITAS.2018.leptons}. 
The different contributions (from pulsars, SNRs and CR interactions) are shown separately. 
The total spectrum exhibits a clear trend toward a decline, shown by the data and well reproduced by the results of our calculations. 
At energies $\gtrsim 10$ TeV, the flux of $e^{+}+e^{-}$ is typically dominated by a few local and recent sources, a fact that reflects in a sensible increase of the fluctuations. 

Moreover, in the right panel of Fig.~\ref{fig:leptons} we show a sample of 20 random lepton spectra from the Monte Carlo computation. 
Note that the spectrum from each realization is rather smooth up to $\sim$~10 TeV, while at higher energies several features appear reflecting the occasional contribution of local and recent sources. 
We discuss more in detail the effect of local sources in Sec.~\ref{sec:locality}.

\begin{figure*}[t]
\centering 
\hspace{\stretch{1}}
\includegraphics[height=0.38\textwidth]{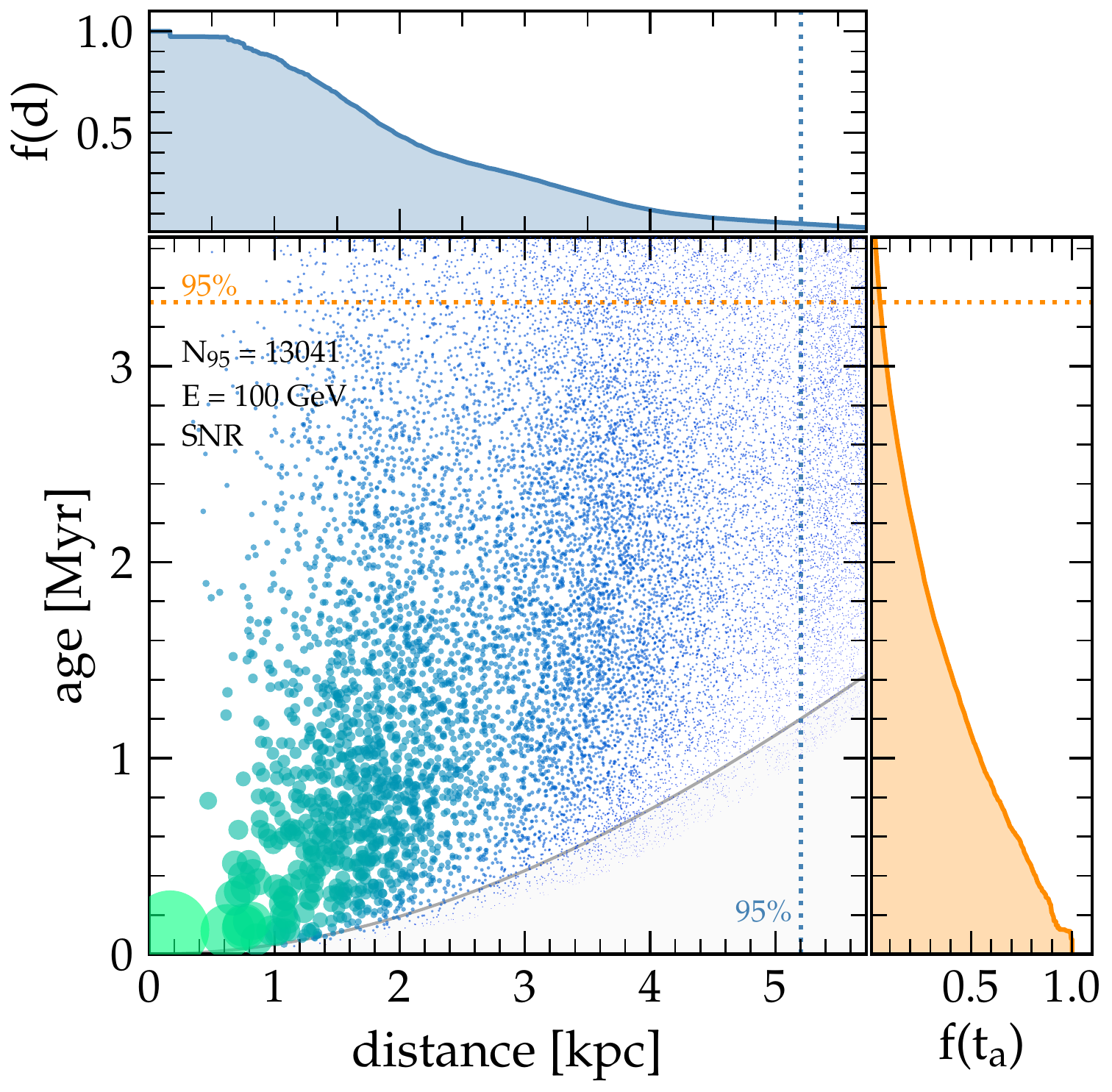}
\hspace{\stretch{1}}
\includegraphics[height=0.38\textwidth]{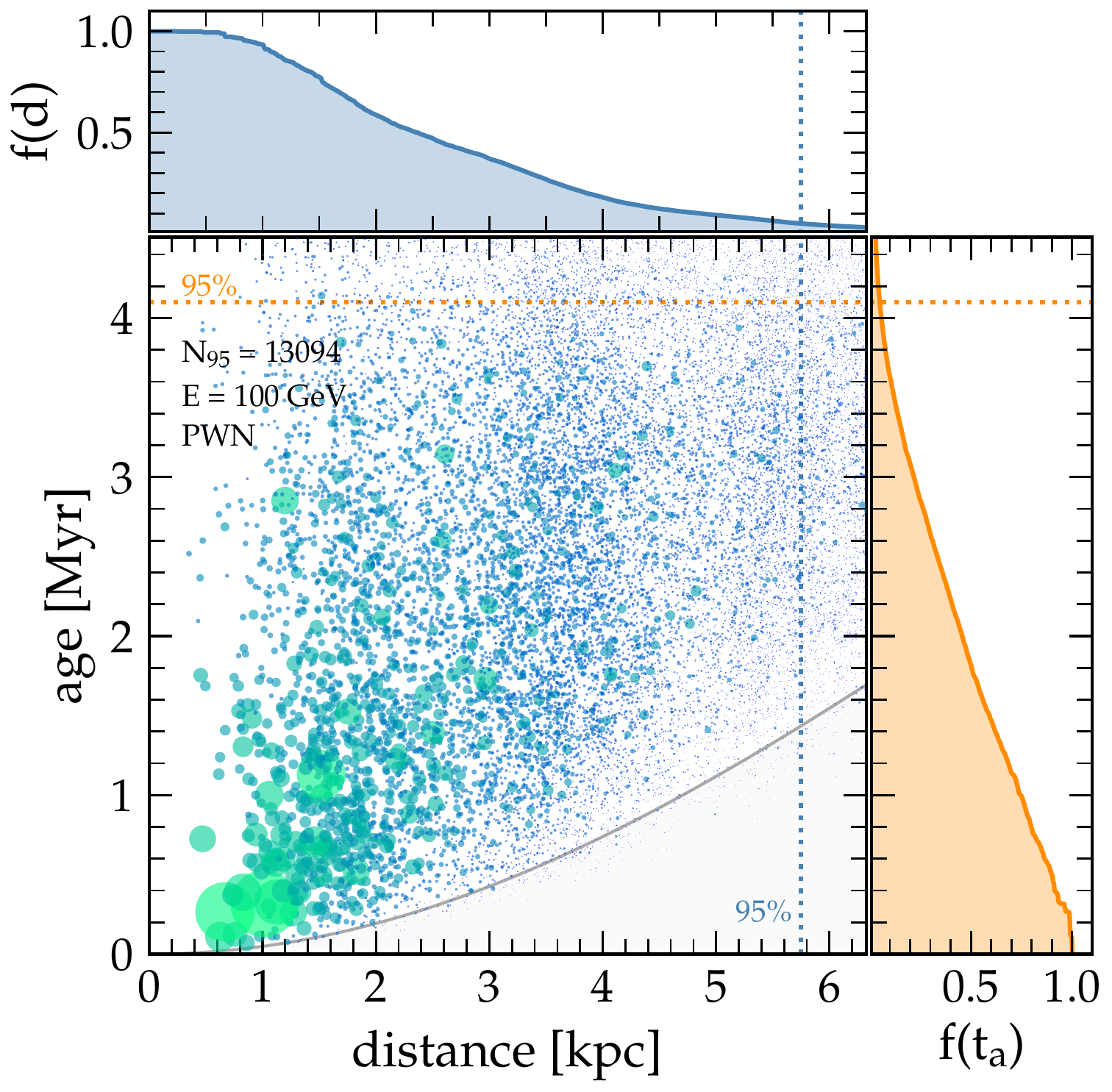}
\hspace{\stretch{1}}

\hspace{\stretch{1}}
\includegraphics[height=0.38\textwidth]{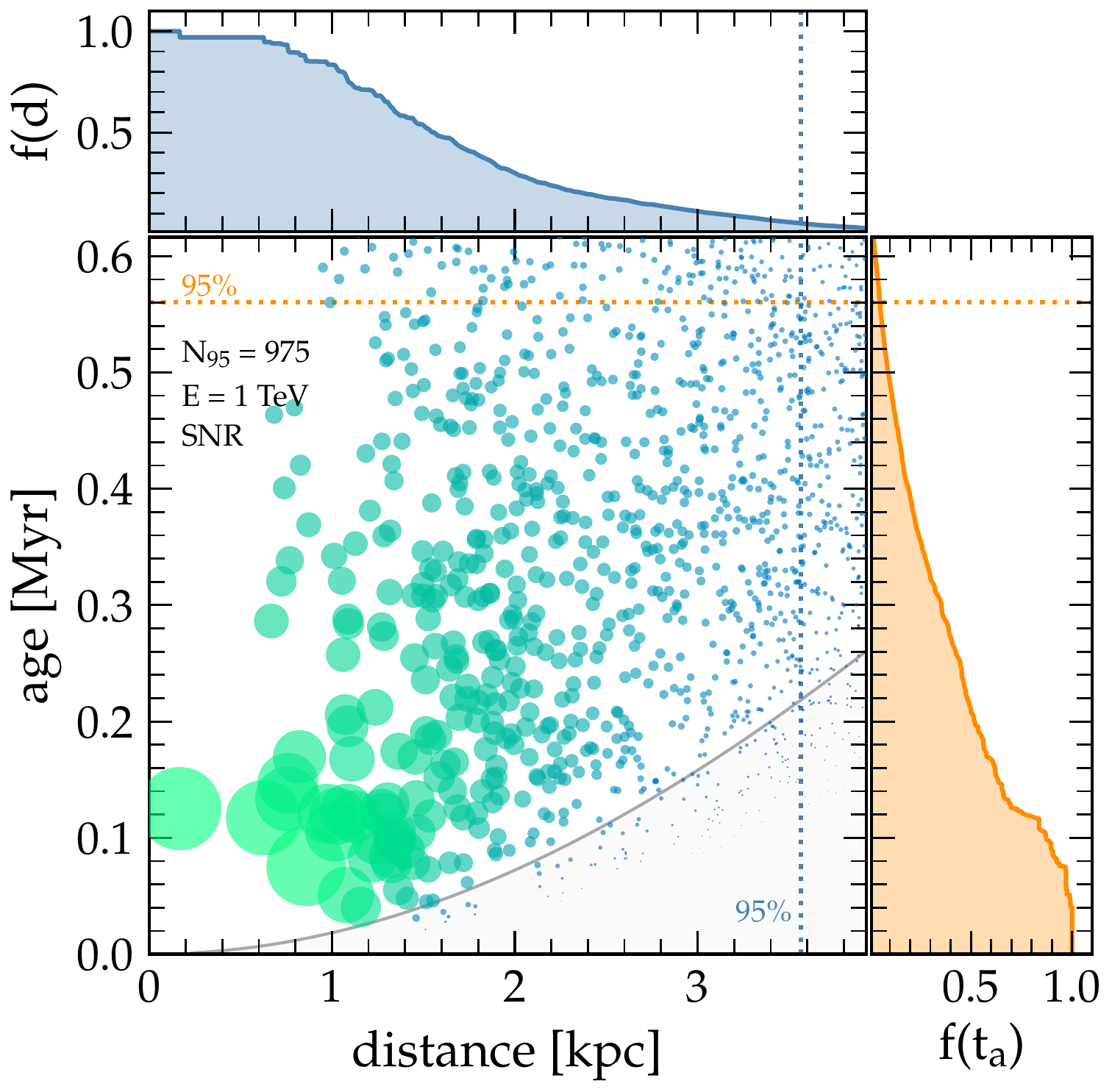}
\hspace{\stretch{1}}
\includegraphics[height=0.38\textwidth]{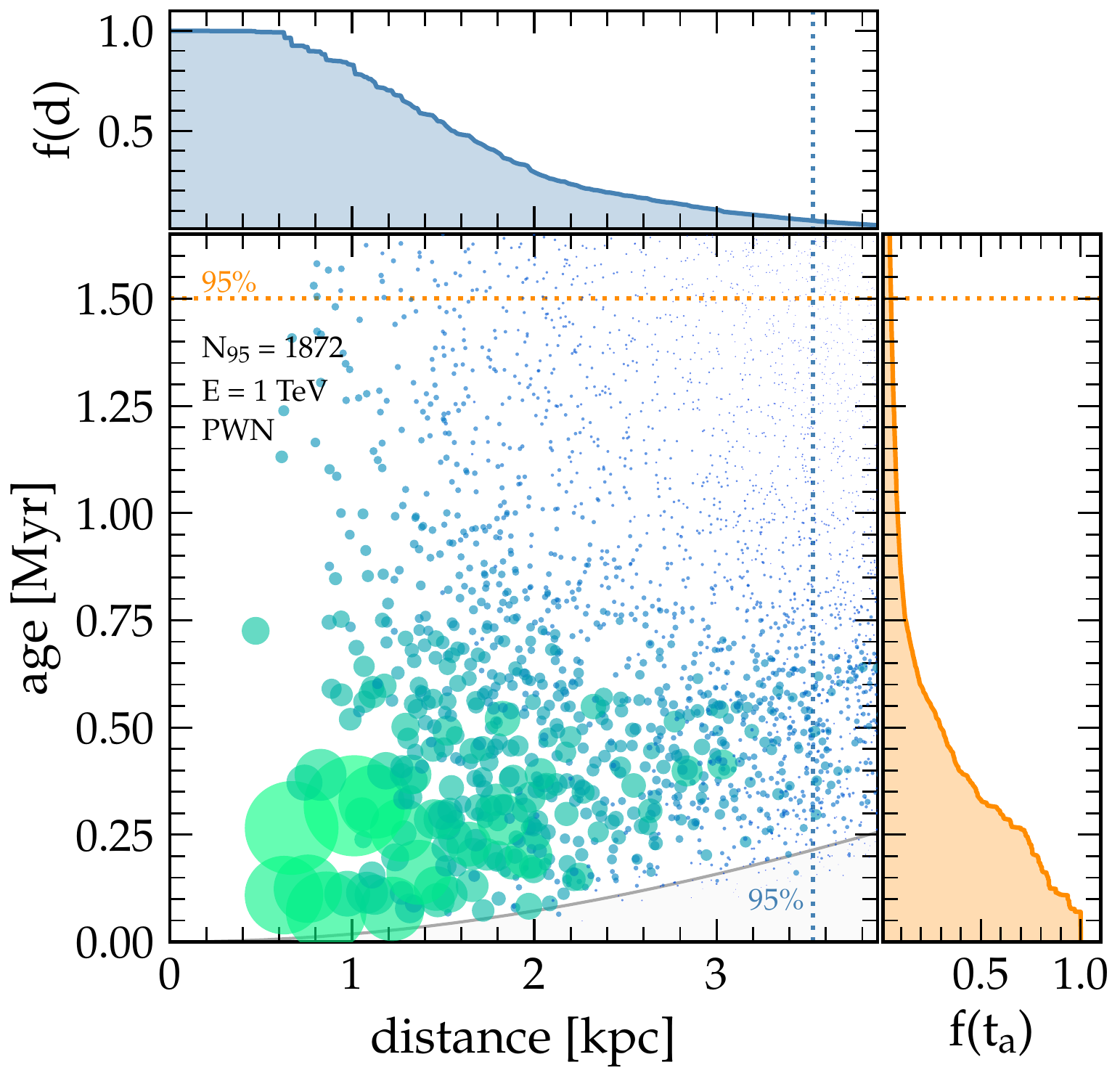}
\hspace{\stretch{1}}

\hspace{\stretch{1}}
\includegraphics[height=0.38\textwidth]{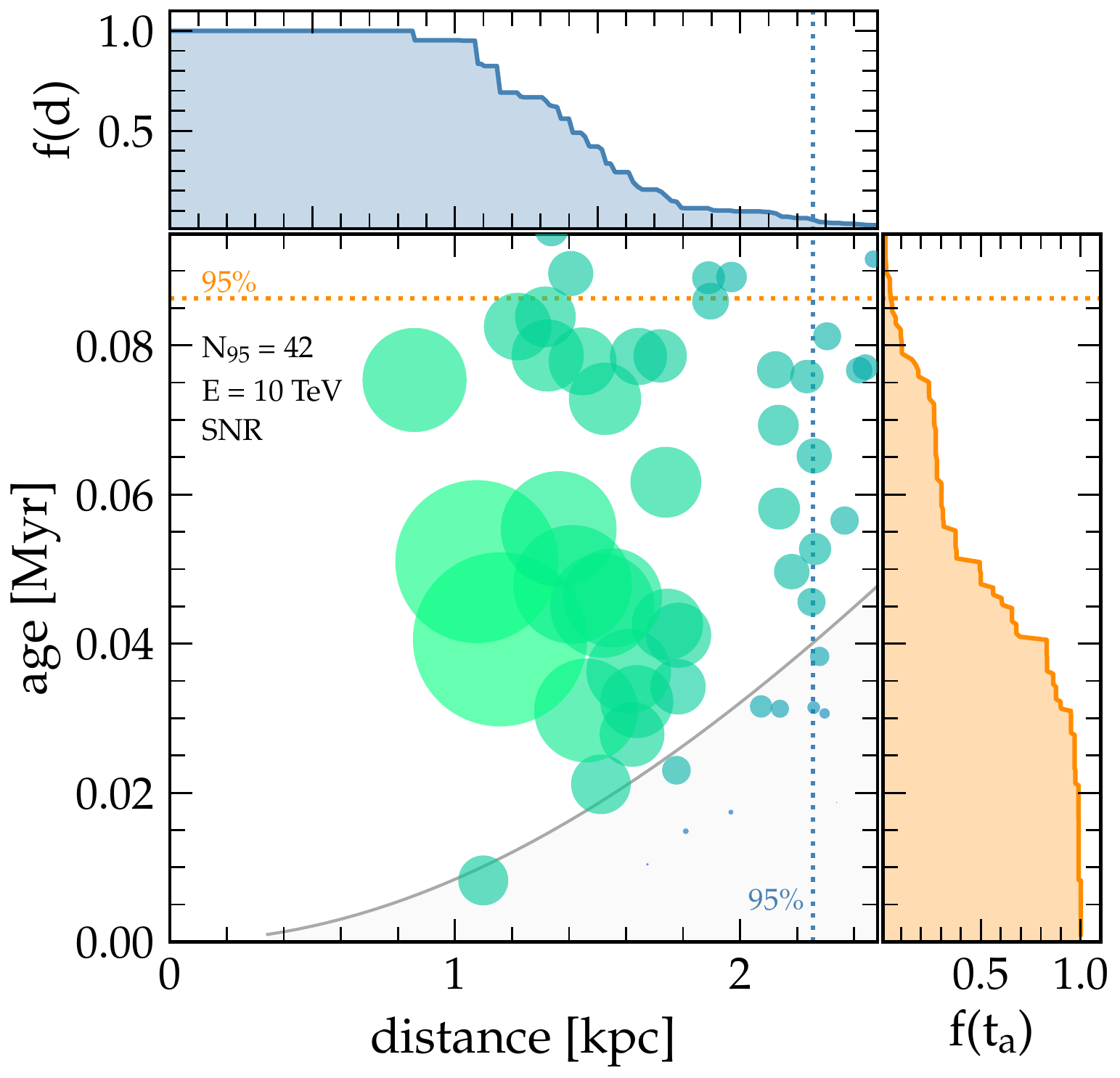}
\hspace{\stretch{1}}
\includegraphics[height=0.38\textwidth]{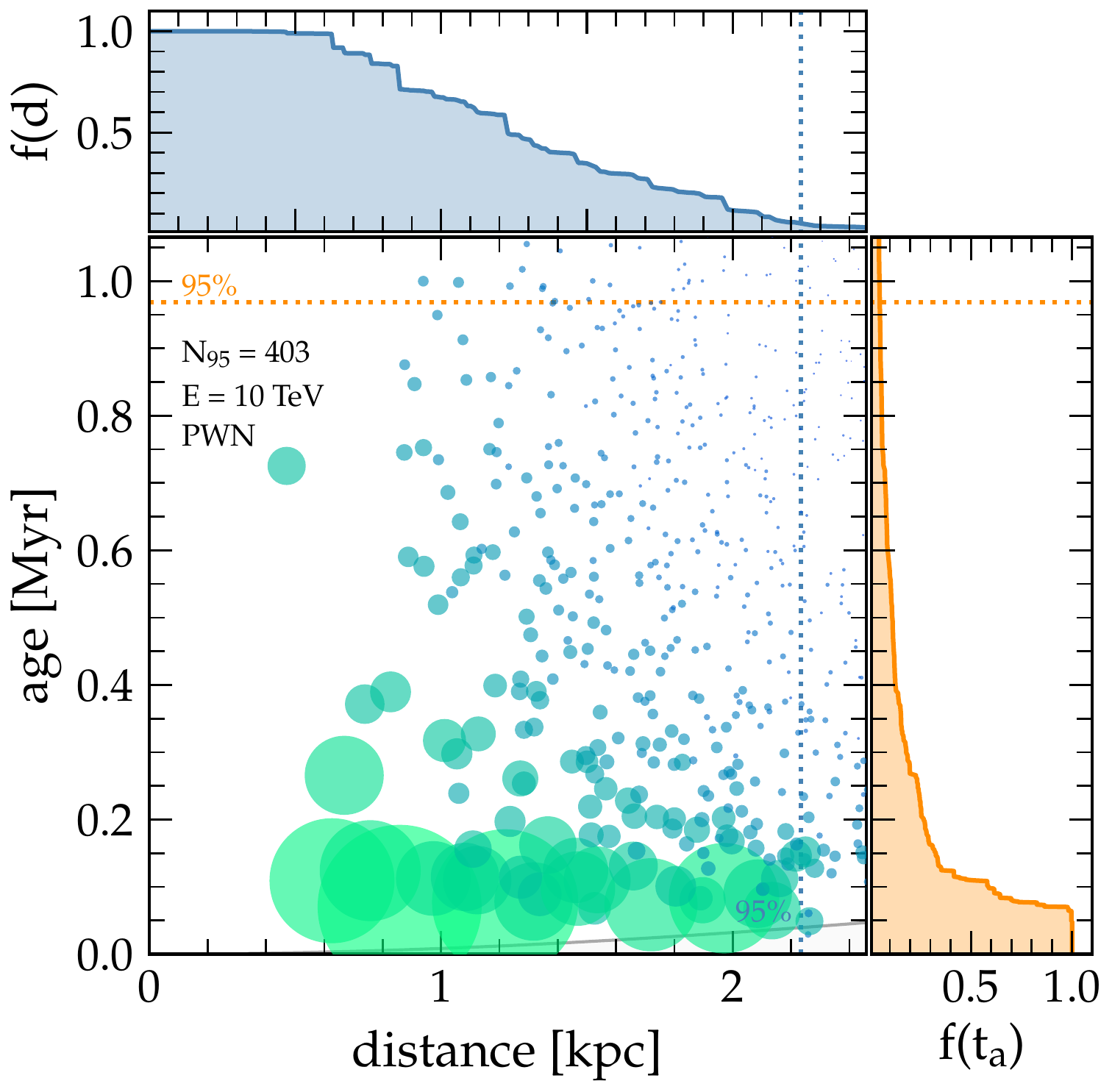}
\hspace{\stretch{1}}

\caption{Temporal and spatial distribution of sources in one realization as a function of energy. The central panel shows the simulated events in the age-distance plane. For illustrative purposes, the size of each symbol is proportional to the logarithm of the relative contribution of that source to the local flux. In the same panel we report (a) the cooling radius $\sim \sqrt{D(E_*) t_a}$ as a function of age and (b) the minimum number of objects required to obtain the 95\% of the observed flux $N_{95}$. In the top panel we show with a blue line the relative contribution to the local fluxes of all the sources at a distance larger than $d$: $f(d)$. Analogously, in the right panel the relative contribution of all the sources with age larger than $t_a$: $f(t_a)$. In both cases a dotted line puts in view where $f$ becomes smaller than 5\%. The left column of panels refer to electron from SNRs, while the right column is for primary positrons (or electrons) from PWNe. The three rows refer to particles of decreasing energy: from top to bottom, the considered energy is 100 GeV, 1 TeV and 10 TeV.}
\label{fig:locality}
\end{figure*}

\subsection{Locality}
\label{sec:locality}

In this section we discuss the effect of local sources on the observed flux at different energies. This issue is of special importance for electrons and positrons due to the fact that their transport on Galactic scales is dominated by radiative losses (for nuclei see, e.g.,~\cite{Taillet2003aa,Evoli2012prd}). At higher and higher energies the number of sources that can contribute to the flux at a given energy becomes smaller and smaller, which in turn implies that fluctuations become increasingly more important. 

This is well illustrated in Fig.~\ref{fig:locality}: the three rows of plots refer, from top to bottom, to energy $100$ GeV, $1$ TeV and $10$ TeV, respectively. 
The left column of plots refer to electrons from SNRs (treated as bursts) and the second to positrons from pulsars (assuming a mean value of the initial rotation period $\langle P_0 \rangle = 100$~ms). Each circle in the plots has a radius proportional to the logarithm of the flux contributed by that source with respect to the total flux. A big circle corresponds to a large flux from that source. For each source the time of appearance and the distance to the Sun are indicated on the $y$ and $x$ axes of the central part of each plot respectively. The dotted vertical and horizontal lines identify the distance and age of the sources below which lie the sources that contribute $95\%$ of the total flux. 

At low energies ($100$ GeV, first row) one can clearly see that a large number of sources contribute to the flux at Earth (in other words, there are many circles with a small radius) and there is not much of a difference in this respect between electrons from SNRs (bursts) and positrons from pulsars (continuous emission). The sources that contribute $95\%$ of the flux are located within $5.2$ kpc from the Sun. 
The situation starts changing for $E=1$ TeV, where radiative losses are more important. We can see that $\sim 10^{3}$ SNRs contribute to the electron flux at this energy, while $\sim 1900$ pulsars contribute to the positron spectrum: the difference is due to the continuous nature of pulsar injection. Also larger size circles start to appear in the small-distance, small-age region (local and recent sources). The sources that contribute $95\%$ of the flux are located within $\sim 3.6$ kpc from the Sun. 
Finally in the $10$ TeV energy bin there is a dominant contribution of few local sources  (large circles). About a hundred sources contribute to the electron flux. As usual a larger number of pulsars contribute to the positron flux because of the different time dependence of their injection rate. Because of this, several of the pulsars with a large contribution to the flux are not necessarily very recent, but they keep contributing to the flux because of their injection extended in time. Most sources contributing to the flux at Earth are within $\sim 2.6$ kpc from the Sun. 

We then compare these numbers with those obtained in Sec.~\ref{sec:spirals}. Here we count the number of sources that are significantly contributing to the local flux, while in Sec.~\ref{sec:spirals} we only provided a rough estimate of the total number of sources that can contribute with a nonvanishing flux, however small it may be. This explains why we obtain here a number of sources which is significantly smaller than the one quoted earlier in the text. We notice that the number of sources becomes of $\mathcal{O}(10)$ at $E \gtrsim 10$~TeV, contrary to previous results that found a number of sources at least one order of magnitude smaller.
As discussed above, this difference is due to the fact that here we use a diffusion coefficient \cite{Evoli2019prd,Evoli2020prd} that provides a good description of the recent AMS-02 data on primary and secondary nuclei (both stable and unstable)~\cite{AMS02.2018.libeb}.

\section{Conclusions}
\label{sec:conclusions}

We presented the results of a calculation of the flux and spectrum of electrons and positrons as produced by SNRs, PWNe and secondary interactions of CRs in the Galaxy, taking into account the stochasticity in the occurrence of SN events, of birth of pulsars in core collapse SN explosions, and the spiral structure of the Galaxy, as recently modeled by~\cite{SteimanCameron2010apj}. The propagation of leptons was treated through a Green function formalism accounting for radiative losses and diffusion, with proper boundary conditions to describe the escape of CRs from the Galaxy. The escape of leptons from PWNe is assumed to occur when pulsars escape the parent remnant (BSN phase), at a time after explosion that depends on the birth kick velocity, which is also extracted at random from the observed distribution of kick velocities. The energy liberated by each BSN depends on the initial spin period $P_{0}$, whose observed statistical distribution is taken into account.

We showed that this model simultaneously reproduces all the available CR lepton data in a very satisfactory way, with perfectly reasonable values of the unknown parameters.

The feature in the electron spectrum, at energies $\gtrsim 40$ GeV, observed by AMS-02 was shown to be the consequence of the transition from Thomson to KN regime of ICS of electrons on the UV component of the background light. We previously presented this result~\cite{Evoli2020prl} using a parametrization of the ICS cross section taken from~\cite{Hooper2017prd}. This parametrization was later proven~\cite{Fang2020arxiv} to provide a rather poor description of ICS losses in the transition region between the two regimes. Here we repeated the calculation using the approach presented by~\cite{Fang2020arxiv} and showed that the feature is still present in the computed electron spectrum, although the required parameters and the UV background are somewhat different from those used in~\cite{Evoli2020prl}, while perfectly compatible with observations.

Our calculations suggest a quite interesting set of conclusions in terms of source spectra: (a) the source spectrum of primary electrons is steeper than that of protons: $\gamma_{\rm e} \sim 2.6$, versus $\gamma_{\rm H} \sim 2.3$, with an efficiency $\sim 0.3$\% assuming the canonical rate of SN explosions in the Galaxy $\mathcal R = 3$ per century; (b) the spectrum of pairs injected by BSNe is required to have slope $1.8(2.8)$ below(above) the break at $E_b = 450$~GeV and an efficiency of $\sim 17$\% for an average initial period $\langle P_0 \rangle = 100$~ms. 

The difference in spectral shape of electrons and protons in SNRs has been invoked several times before, and is clearly at odds with the expectation based on standard DSA. It has been proposed that perhaps radiative losses of electrons in SNRs before their release in the ISM may account for this difference~\cite{Diesing2019prl}, although it is not clear whether the required conditions are generic. 
In the case of pairs produced in BSNe, the available multifrequency investigations~\cite{Kargaltsev2008aipc} suggest that the source spectra should be harder than what inferred by the present investigation by about $\sim 0.3$.
In both cases, the spectra are somewhat steeper than expected if one assumed a universal injection for electrons and protons in SNRs and if compared with gamma-ray observations of PWNe and BSNe~\cite{Kargaltsev2008aipc}. These differences however might suggest that additional effects take place in the source surroundings.

Aside from the shape of the electron and positron spectra injected in the Galaxy, another important conclusion we reached concerns its extent in energy: we showed that the spectral steepening observed at $\sim$TeV can be reproduced (within the systematic uncertainties) as the result of the cut-off in the injection spectra and transport through the Galaxy.

Finally, a result of fundamental importance relates to the number of contributing sources at different energies and cosmic variance. Our findings are shown to be very little affected by the exact distribution of sources: the number of contributing sources is much larger than in previous works at all energies, and the fluctuations associated with different distributions of sources are correspondingly smaller. This is a direct consequence of the adopted diffusion model, driven by the recent AMS-02 data on primary and secondary nuclei (both stable and unstable)~\cite{Evoli2019prd,Evoli2020prd}. The difference with respect to previous findings is particularly impressive at high energies: at 1 TeV, where the dropoff in the lepton spectrum is detected, about 10$^3$ sources are found to contribute to the local flux, with their number decreasing to $\sim10$ only at $\sim 10$ TeV. In summary, for the propagation scenario that best fits all the available AMS-02 data simultaneously, the possibility that just one source is responsible for the measured fluxes of leptons at Earth at the highest energies appears to be ruled out.

\begin{acknowledgments}
C.E. is very grateful to Kathrin Egberts for helpful comments on HESS CR measurements.
We acknowledge support by INAF and ASI through Grants No.~INAF-MAINSTREAM 2018 and ASI/INAF No.~2017-14- H.O and by the National Science Foundation under Grant No.~NSF PHY-1748958.
We acknowledge the use of the CRDB~\cite{Maurin2020crdb} and ASI~\cite{DiFelice2017asi} databases for providing CR measurements.
\end{acknowledgments}

\bibliographystyle{myapsrev4-2}
\bibliography{2020-positron-pwne.bib}

\end{document}